\documentstyle[12pt,aps,epsfig,psfig19,floats]{revtex}

\newcommand{\mb}[1]{\mbox{\boldmath$#1$}}

\renewcommand {\baselinestretch}{1.}
\def\begineq{\begin{equation}}
\def\endeq{\end{equation}}
\def\la{\left<}
\def\ra{\right>}

\renewcommand{\theequation}{\arabic{section}.\arabic{equation}}
\newcommand{\aleq}{\mbox{\ 
\raisebox{-.9ex}{$\stackrel{\textstyle<}{\sim}$}\ }}
\newcommand{\ageq}{\mbox{\
\raisebox{-.9ex}{$\stackrel{\textstyle >}{\sim}$}\ }}

\begin{document}
\bibliographystyle{prsty}
\title{Phase Diagrams for Sonoluminescing Bubbles}
\author{
Sascha Hilgenfeldt$^1$, Detlef Lohse$^1$, and Michael P. Brenner$^2$}
\address{
$^1$Fachbereich Physik der Universit\"at Marburg,\\
Renthof 6, 35032 Marburg, Germany \\
$^2$
Department of Mathematics, MIT,\\ Cambridge, MA 
02139}
\date{\today}

\maketitle

\begin{abstract}
Sound driven gas bubbles in water can emit light pulses.
This phenomenon is called sonoluminescence (SL).
Two different phases of single bubble SL are known:
diffusively stable and diffusively unstable SL.
We present phase diagrams in the gas concentration vs forcing
pressure state space and also in the ambient radius vs gas
concentration and vs forcing pressure state spaces.
These phase diagrams are based on the thresholds for energy focusing
in the bubble 
and two kinds of instabilities, namely (i) shape
instabilities and (ii) diffusive
instabilities.
Stable SL 
only occurs in a tiny parameter window of large forcing
pressure amplitude $P_a
\sim 1.2 - 1.5$atm and low gas concentration of less than
$0.4\%$ of the
saturation. The upper concentration threshold becomes smaller
with increasing forcing.
Our results quantitatively agree with experimental results
of Putterman's UCLA group 
on argon, but not on air.
However, air bubbles and other gas mixtures can also
successfully be treated in this
approach if in addition (iii) chemical instabilities are considered. 
--
It is only the Rayleigh-Plesset ODE approximation of the bubble dynamics
--  extended in an adiabatic approximation to include
mass diffusion effects  -- 
which allows us to explore 
considerable portions of parameter space.
Therefore, we checked the adiabatic approximation by comparison
with full
numerical solution of the advection diffusion PDE and find good agreement.
\end{abstract}


%-------------------------------------------------------

\newpage

\section{Introduction}\label{intro}
\setcounter{equation}{0}

\subsection{The phenomenon}
\noindent
A gas
bubble levitated in a strong acoustic field
\begineq
P(t) = P_a \cos \omega t
\label{poft}
\endeq
can emit bursts of
light so intense as to be observable by the naked eye
\cite{gai90,bar91,bar94,hil94,bar95,loe93,loe95,hol94}. 
Here, $P_a$ is the forcing pressure amplitude
and $\omega/2\pi$ the 
frequency of the
forcing field. 
This
phenomenon is called single bubble sonoluminescence 
(SL).
The light pulse is shorter than 50ps \cite{bar91,mor95}.
Precise
experiments by Putterman's group at UCLA
\cite{bar91,bar94,hil94,bar95,loe93,loe95} have 
revealed many
surprising and intriguing properties of sonoluminescing 
bubbles.
SL only occurs in a narrow parameter range.
The adjustable
experimental parameters we focus on here are the forcing 
pressure amplitude
$P_a$ and the gas concentration $c_\infty$ far from the 
bubble. 
Single bubble SL is found only for
large (compared to the ambient pressure $P_0$) forcing 
pressure $P_a \sim
1.2 - 1.5$atm and small (compared to the saturation 
$c_0$) gas concentration
$c_\infty$.

We report the results on argon bubbles first.
Two distinct phases of single bubble SL are known 
\cite{bar95}:
(i) {\it Unstable SL} occurs
in the concentration range 
$c_\infty /
c_0 = 6\% - 26\%$.
This phase is characterized by an 
increase of the relative phase of light emission with respect to the
driving pressure
on the slow  diffusive time scale 
$\sim 1s$,
followed by a rapid breakdown and another subsequent 
increase.
The light intensity itself behaves in the same way and the bubble
is reported to be dancing or jiggling \cite{gai90,bar95}.
This state of SL is also unstable in the sense that
often 
all of a sudden the
bubble dies. 
(ii) {\it Stable SL} occurs in argon bubbles
at very low gas concentrations $c_\infty/c_0\sim 0.4\%$. 
The diffusively stable state is characterized
by the constancy of
the relative phase
of light emission over billions of cycles (days).
The same is true for the light intensity.

For air bubbles 
the same two phases exist, however, for gas concentrations about two orders of
magnitude larger than for argon bubbles: Stable SL is observed for
$c^{air}_\infty/c_0 \sim 10-20\%$ \cite{bar95}, unstable SL for even higher
concentrations.

Another controllable
parameter is the temperature of the liquid.
Upon decreasing the temperature of the ambient water 
from
room temperature to slightly above freezing, the light 
intensity increases by
two orders of magnitude \cite{bar94}.
Abrupt transitions in the light intensity
with the liquid temperature are found for SL in nonaqueous
fluids \cite{wen95}.
Using different fluids (but the same gas species) also results in great
differences in SL intensity \cite{hil95,wen95}.

\subsection{A hydrodynamic approach}
The goal of this paper is to figure out which features of SL can be
accounted for in a purely hydrodynamic approach
\cite{loe93} and to which extent
they may reflect other, non-hydrodynamic effects, e.g. chemistry.

\begin{figure}[htb]
\setlength{\unitlength}{1.0cm}
\begin{center}
\begin{picture}(11,10)
\put(-0.6,0.0){\psfig{figure=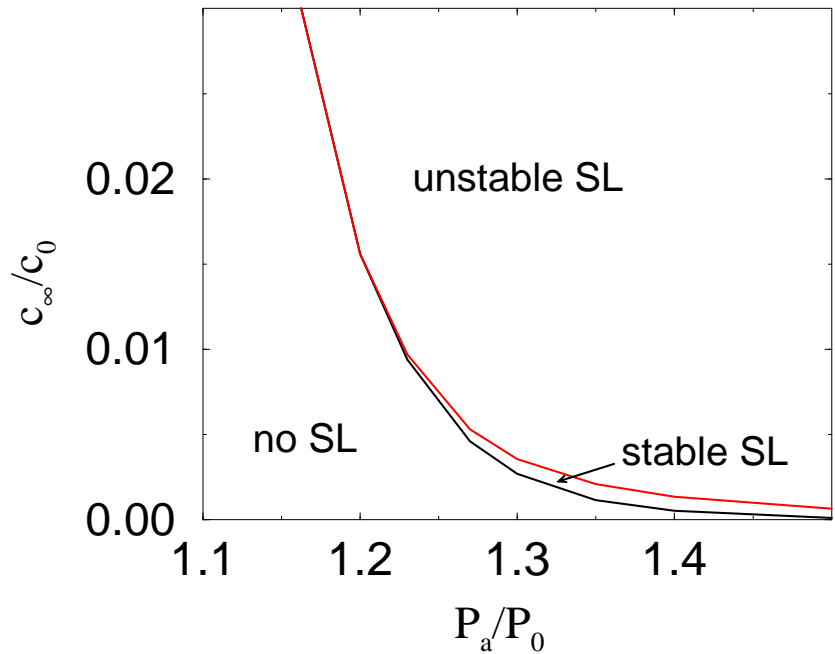,width=12cm,angle=-90.}}
%\put(-0.6,0.0){\psfig{figure=dummy.eps,width=12cm}}
\end{picture}
\end{center}
\caption[]{ 
Phase diagram in the $c_\infty/c_0$ vs $P_a/P_0$ parameter space.
The three phases represent stable SL, unstable SL, and no SL.
For lucidity we do not draw the upper and the
right borderline of the unstable
SL regime towards a no SL regime as they are less precisely defined. 
}
\label{phase_dia}
\end{figure}

Our main result is the phase diagram figure 
\ref{phase_dia} in the 
$P_a -c_\infty$ phase space.
It is obtained from hydrodynamic calculations of the bubble dynamics and
the fluid dynamical and diffusive processes outside the bubble.
For  given forcing pressure amplitude $P_a$ and gas concentration 
$c_\infty$ we
predict with this diagram whether the bubble is in the diffusively 
unstable SL state,
the diffusively stable SL state, or in no SL state at 
all.

Besides the forcing pressure amplitude $P_a$ and the gas concentration
$c_\infty$,
two further parameters have to be addressed here: The 
frequency $\omega/2\pi$ of
the forcing field and the ambient radius $R_0$ of the 
bubble, i.e., the bubble radius at ambient normal conditions
of $P_0=1$atm and 293K. The frequency 
is set to a fixed value so that the forcing acoustic field
(\ref{poft})  corresponds to a resonance of the container, in
order that the 
bubble remains trapped in a pressure antinode.
All of the analysis corresponds to $\omega = 2\pi \cdot 
26.5kHz$ as
applied in Barber et al.'s experiment \cite{bar95}; the period of
the forcing field is thus $T= 2\pi / \omega = 38 \mu s$.

The ambient radius $R_0$ of the bubble is not an 
adjustable parameter but the
system chooses $R_0$ dynamically,
as emphasized already by Putterman. The approach followed here includes 
this (diffusive) dynamics, so that we can predict the ultimate ambient radius.
Our results are consistent with
Mie scattering radius measurements \cite{bar95}. 

What are the necessary requirements for SL to occur?
First, the bubble has to be stable towards shape 
oscillations 
\cite{ple54,bir54,ell70,str71,pro77,bre95}. We have 
identified
three types of shape instability:
Two instabilities of a parametric type 
acting on  relatively slow time scales of $\sim 1-100\mu s$.
As discussed below, these instabilities are quite gentle
and the bubble can survive them by 
pinching off micro bubbles.
This pinch off causes the aforementioned
break in the relative phase of light
emission in the unstable SL state \cite{bar95}. The
third type of shape instability however acts on a very
short time scale $\aleq 10^{-9}s$. It is
so violent that it ejects the bubble from the trapping sound field.
We call it the  Rayleigh-Taylor instability as it occurs when
gas from inside the bubble is accelerated towards the fluid.

What is the energy focusing process in the bubble?
Many speculations abound in literature
\cite{sch92,fli89}.
Jarman \cite{jar60} (for multi bubble SL) and later 
Greenspan and Nadim \cite{gre93} and Wu and Roberts 
\cite{wu93} suggest that {\it shocks }
detach during the compression of the bubble and focus
to the center of the bubble, thereby compressing the gas 
so strongly
that light can be emitted, either by ionization and 
subsequent bremsstrahlung \cite{wu93,loe93}  or
by black body radiation \cite{hil95}.
 In ref.\ \cite{bre96d} we suggest an alternate energy focusing
 mechanism. 
The idea is that the bubble acts as a driven acoustic resonator
which switches on
when the damping 
losses through viscosity and acoustic radiation are 
smaller than the energy input during the collapse.
The acoustic
energy accumulates and  finally results in light pulses. 
The above mentioned
abrupt transition in the SL intensity with increasing temperature
\cite{wen95} can be accounted for within our theory,
as well as the 
dependence of the light intensity on different liquids and gases.
We refer to ref.\ \cite{bre96d} for
a detailed discussion.

In this paper we
take as 
criterion for energy focusing and
the resulting light production that 
the Mach number of the bubble wall has to be 
larger than one
\cite{ll87}. This criterion corresponds to the onset
of SL within 
conventional shock theories \cite{jar60,wu93,gre93} and within  
our alternate energy focusing mechanism \cite{bre96d}.
Together with the requirement of bubble stability, it gives the
boundary of the stable SL regime of Fig.\ \ref{phase_dia}. 
For argon bubbles
the narrow parameter range where stable SL exists
is in agreement with experiments of Barber et al.
\cite{bar95}.  A qualitative argument for this agreement was
previously given by L\"ofstedt et al. \cite{loe95}.

\subsection{Chemical instabilities}
For air bubbles there are severe deviations  between the 
hydrodynamically calculated  phase
space diagram and the experimental
measurements \cite{bar95}.
The parameter regime where stable air bubbles
should exist is close to that of argon;
however, experimentally stable
SL is found for gas concentrations as 
large as
$c^{air}_\infty / c_0 \sim 20\%$ \cite{bar95}.
Because of this discrepancy, L\"ofstedt et al. hypothesize
\cite{loe95} a ``yet unidentified mass ejection mechanism''. In ref.\
\cite{loh96} we have suggested that this mechanism is chemical. Indeed,
when considering besides (i) shape instabilities and (ii) diffusive
instabilities also (iii) chemical instabilities, our results can be
extended to gas mixtures and are then in quantitative agreement with
the UCLA experiments, as shown in detail in ref.\ \cite{bre96b}.
The idea is that
because of the high temperatures achieved in the bubble
nitrogen and/or oxygen is destroyed
and reacts to NO$_3^-$, NO$_2^-$, and/or NH$_4^+$ and only pure
argon remains in the bubble. Thus for air which contains about $1\%$
argon the gas concentrations $c_\infty^{air}$
in water have to be about two orders
of magnitude higher than for pure argon. The central parameter
is thus the argon (or inert gas)
 concentration $c_\infty^{Ar} = q c_\infty^{mixture}$
 in the dissolved gas. Here, $q$ is the percentage of argon in the
 mixture; for air $q=0.01=1\%$.
The nitrogen dissociation theory suggests that
when 
adjusting $q$ properly,
no degassing is necessary any more \cite{loh96,bre96b}.

In this paper we work out the basics of
our hydrodynamic approach and
restrict ourselves to pure argon bubbles for which no chemical instabilities
(i.e., reactions) can occur. However, by considering the
chemical instabilities properly \cite{loh96,bre96b}, our results here
can directly be extended to any gas mixture and are found to agree
with the UCLA experiments.

\subsection{Necessary approximations}
How can we examine the huge multi--dimensional parameter space ($P_a$, 
$c_\infty$, and $R_0$)?
Given that the dynamics involves time scales spanning 
eleven orders of
magnitude
(from the time scale of the light flash ($< 50 ps$)
to the diffusive time scale ($\sim 1s$)),
it is necessary to make approximations
in modeling the hydrodynamics of sonoluminescence.
The full hydrodynamic problem involves solving the  3D
Navier-Stokes equation both inside and outside the 
bubble,  
coupled with equations of heat transfer and
gas transfer,
accompanied by the correct boundary
conditions at the interface and at infinity and the
equations of state. Moreover, at least in principle
the radiation fields 
need to be coupled to the fluid.  This set of equations
must be studied not only as a function of parameters but also
over millions of oscillation periods
of the bubble; it should be emphasized that
the relevant question for 
sonoluminescence experiments
is not the transient that occurs for the first few 
cycles but
rather the nature of the long time limit.
This complete formulation is both computationally and 
theoretically
intractable.  In order to make progress, the problem 
must be simplified.
To date, two different avenues have been pursued:

The first approximation was proposed long ago by Lord 
Rayleigh
\cite{ray17} and elaborated
upon by Plesset \cite{ple49}, Taylor \cite{tay50}, 
Lauterborn \cite{lau76}, Prosperetti \cite{ple77,pro77}
and others \cite{kel80}, in the context of studies of
cavitation. 
The idea is to consider
the bubble as a perfectly spherical cavity, with
the pressure inside the bubble having no spatial 
variations.  The temporal
variation of the pressure follows from an equation of 
state.
In this approach the full dynamics is reduced to the 
Rayleigh-Plesset
ODE \cite{bre95b}.
This formulation allows very long time calculations 
of the bubble dynamics, but it completely ignores 
the dynamics inside
the bubble producing the light.
Later on, Plesset \cite{ple54}, Strube \cite{str71}, and 
Prosperetti
\cite{pro77} extended
this type of approach to deal with {\it shape oscillations}
while Epstein and Plesset \cite{eps50}, Eller and Crum \cite{ell69,ell70},
Crum and Cordry \cite{cru94} and finally 
Fyrillas and Szeri \cite{fyr94} and 
L\"ofstedt et al. \cite{loe95}
included {\it diffusive effects.}
We call this approach the {\it RP-SL-bubble approach}.
Clearly, shock formation \cite{jar60,gre93} or the building up 
of the acoustic waves \cite{bre96d} inside the bubble
will modify the dynamics of $R(t)$. But it is our belief
that the results of this paper are robust towards these changes
and 
it is only within this RP-SL bubble approach that the exploration of
the full SL parameter space and the calculation of
phase diagrams are currently manageable.
Full numerical simulations as in \cite{mos94,vuo95}
are by far numerically too expensive
to do such an analysis. 

The second type of approximation traditionally
made is complementary to the 
first, and
focuses on the interior of the bubble and
above mentioned shock formation processes
\cite{gre93,wu93,mos94,szeri}.
The
spherically symmetric gas dynamics equations are solved 
inside the bubble
and coupled to the Rayleigh-Plesset equation.
Simplifications are typically employed in modeling the 
gas dynamics,
for example neglecting heat and viscous
dissipation.
These calculations can only be carried out for a few 
oscillation
periods, and thus are not able to resolve cumulative 
effects building
up over many oscillations.

\subsection{The Rayleigh-Plesset equation}
The Rayleigh--Plesset (RP) equation 
\cite{loe93,ray17,lau76}, on which
the entire analysis of this paper is based,
describes the
dynamics of the bubble radius,
\begin{eqnarray}
R \ddot R + {3\over 2} \dot R^2  &=&
{1\over \rho_w} \left(p(R,t) - P(t) - P_0 \right)
             \nonumber \\
	            &+& {R\over \rho_w c_w} {d\over dt}
		    \left( p(R,t) - P(t)\right) - 4 \nu 
{\dot R \over R} -
		    {2\sigma \over
		    \rho_w R}.
		    \label{rp}
\end{eqnarray}
Typical parameters for an argon bubble in water are
the surface tension
$\sigma = 0.073 kg/s^2$, the water viscosity 
$\nu = 10^{-6} m^2/s$, density 
$\rho_w= 1000 kg/m^3$, and
speed of sound $c_w=1481 m/s$.
The driving frequency of the acoustic field is
$\omega/ 2\pi = 26.5 kHz$
and the external pressure 
$P_0= 1$atm.
We assume that 
the pressure inside the bubble
varies according to
\begineq
p(R(t)) = P_0 \left( {R_0^3 - h^3\over R^3(t) - h^3 
}\right)^\gamma 
\label{pressure}
\endeq
where
$h= R_0/8.86$
is the hard core van der Waals radius.
The exponent $\gamma$ is the effective polytropic 
exponent of the gas.
Plesset and Prosperetti \cite{ple77} calculated how it depends on 
the (thermal)
Peclet number
$Pe = R_0^2\omega/\kappa$ which gives the ratio between 
the bubble length scale $R_0$
(which we take as $\approx 5\mu m$ for the estimates in this paragraph)
and the thermal diffusion length 
$\sqrt{\kappa / \omega}$.
The thermal diffusivity $\kappa$ for argon is
$\kappa\approx 2\cdot10^{-5}m^2/s$,
which yields $Pe\approx 0.2$ and according to fig.\ 1 
of \cite{ple77}, the effective polytropic exponent $\gamma=1$.
As discussed below, the RP equation 
contains much smaller time
scales than $\omega^{-1}$. One could therefore argue that these smaller 
time scales may enter
into the calculation of $Pe$,  so that
the
frequency $\omega$ should be replaced by $|\dot R|/R$.
This estimate would lead to
$Pe(t)$ as large as $10^4$ at instants of rapid bubble wall
movement which  implies
$\gamma \approx
5/3$ for argon. However, since $Pe (t) \gg 1$ only holds in 
very small time
intervals $\sim 1ns$, the global dynamics  are
not affected by setting
the effective polytropic  exponent $\gamma = 1$ uniformly
in time. Note 
that with $\gamma = 1$
eq.\ (\ref{pressure}) should not be thought of as 
equation of state but rather
as a process
equation parametrizing the isothermal conditions at the 
bubble wall,
induced by the large heat capacity of water.
The choice of $\gamma=1$ is
confirmed by the full numerical simulations of Vuong and 
Szeri \cite{vuo95}
and by the approximation of 
Kamath et al.\ \cite{kam93}. 
Note that, as a 
consequence, there are heat
fluxes back and forth across the bubble wall.

\begin{figure}[thb]
\setlength{\unitlength}{1.0cm}
\begin{center}
\begin{picture}(11,10)
%\put(0.5,7.5){\LARGE a)}
%\put(0.5,5){\LARGE b)}
\put(0.0,0.0){\psfig{figure=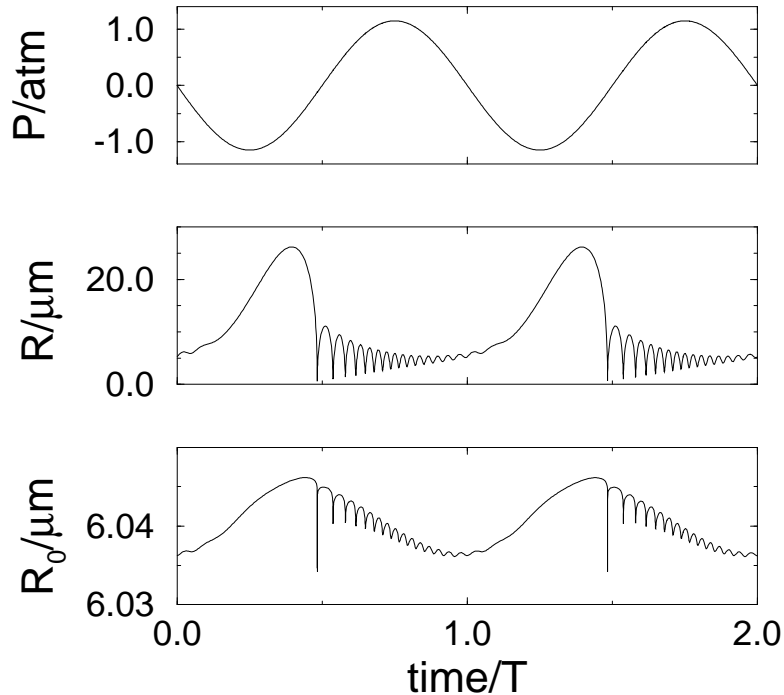,width=12cm,angle=-90.}}
%\put(0.5,0.5){\psfig{figure=dummy.eps,width=12cm}}
\end{picture}
\end{center}
\caption[]{ (a) Forcing pressure 
$P(t) = P_a \cos \omega t $, $P_a=1.15$atm for two cycles and
the corresponding (b) $R(t)$ and (c) $R_0(t)$.
The bubble is near an
equilibrium state. The gas concentration is 
$c_\infty / c_0 = 0.035$.
}
\label{roft}
\end{figure}

The radius $R(t)$ corresponding to the forcing pressure 
(\ref{poft})
with $P_a=1.15$atm is shown in figure \ref{roft}b. Four time scales are hidden 
in the RP equation: The period $T=38\mu s$ of the 
external forcing $P(t)$ (figure \ref{roft}a), the 
intrinsic frequency $ \sqrt{3\gamma P_0/(\rho_w 
R_0^2)}/2\pi
\approx (1.8\mu s)^{-1}$ of the oscillating bubble which is 
the
frequency of the afterbounces, the time scale of viscous 
damping
$R_0^2/\nu \approx 25\mu s$, and the duration $\sim 0.1-1ns$ of 
the bubble
collapse, estimated in \cite{loe93}. 
A fifth time scale, determined by the surface tension 
$\sqrt{R_0^3\rho_w/
\sigma }\approx 1.3\mu s$ 
is only important for large $\sigma$ or small $R$; for typical $R_0$,
it only slightly changes the intrinsic
time scale which is of the same order of magnitude. 

Also {\it diffusive processes} can  be understood within the
RP approach to the SL bubble \cite{loe95,fyr94,bre96}.
At first sight
this is surprising because there is no diffusive time scale in the RP
equation.
For a qualitative understanding why this works nevertheless look at
the bubble radius $R(t)$ (figure \ref{roft}b), resulting from (\ref{rp}).
For large $R(t)$ the pressure
inside the bubble will be low and gas diffuses into the bubble (rectified
diffusion). For small $R(t)$, on the other hand, the bubble will shrink because
of the enormous pressure inside \cite{ell64}.
This concept was made quantitative by Fyrillas and Szeri \cite{fyr94}
and L\"ofstedt et al.\ \cite{loe95}. 
The main idea is a separation of the slow diffusive time scale from
all time scales in (\ref{rp}).
We therefore call this approach the {\it adiabatic} approximation
of diffusion.
As we will see the balance between growth and shrinking
is very delicate. In fig.\ \ref{roft}c we present the ambient radius
(calculated in section \ref{compare})
$R_0(t)$ which represents the mass $m=4\pi
R_0^3\rho_0/3$ of the bubble, with the ambient density $\rho_0\approx 1.6kg/m^3$
for argon.
It corresponds to the $R(t)$ curve in fig.\
\ref{roft}b. The exchange processes between the bubble and the liquid can be
very violent. In fig.\ \ref{roft}c the bubble's mass increases to $100.5\%$
of its initial value and decreases to $99.8\%$
thereof while after one full cycle it again takes its initial value.
These are representative values for
argon bubbles near the onset of  the SL regime; for larger forcing,
the exchange processes become even more violent.

\subsection{Organization of the paper}
The paper is organized as follows.
In section II we analyze the
bubble stability
with respect to shape oscillations.
We then
give as necessary criterion for light emission energy focusing
in the bubble, either through a shock or through
acoustic resonance
(section \ref{shocks}).
Section \ref{diffstab} constitutes the main part of the paper.
We calculate
the diffusive instability with Fyrillas and 
Szeri's \cite{fyr94} and 
L\"ofstedt et al.'s \cite{loe93} adiabatic approximation.
The main result are  phase diagrams
in the $c_\infty - P_a$, the $R_0-P_a$, and the $R_0 - 
c_\infty$
parameter spaces. Only in a very small parameter domain 
does stable SL
occur.
To analyze unstable SL 
we calculate growth rates and compare them to
Putterman's measurements of diffusively unstable SL argon bubbles 
\cite{bar95}. 
Concentration profiles from a full numerical solution of 
the advection diffusion PDE are
presented in section \ref{compare}, where
we also check the validity of the
adiabatic approximation by comparison with full numerical solutions of
diffusive dynamics. We find 
experimentally undetectable discrepancies which vanish 
in the Schmidt number
$Sc\to \infty $ limit.
Section \ref{concl} presents conclusions.

\section{Shape stability}\label{shapestab}
\setcounter{equation}{0}
For sonoluminescence to occur and for the bubble to 
remain
oscillating for billions of cycles, the bubble must be 
stable to shape
oscillations.  First, following the pioneering work of 
Plesset \cite{ple54},
Strube \cite{str71}, 
and Prosperetti \cite{pro77}, we derive equations for 
the deviations
of the bubble from a spherical shape, and then proceed 
to analyze them.

\subsection{Dynamical equations}
We focus on the stability of the radial solution
$R(t)$.
Consider a small distortion
of the spherical interface $R(t)$,
$$R(t) + a_n(t) Y_n(\theta , \phi ),$$
where $Y_n$ is a spherical harmonic of degree
$n$.  
The goal is to determine the dynamics $a_n(t)$ for each 
mode. 
Plesset's \cite{ple54} derivation
follows the same spirit as the derivation of the 
Rayleigh-Plesset
equation.
The potential flow outside the bubble is constructed
to satisfy the boundary condition that the velocity at 
the bubble
wall is $\dot{R} + \dot{a}_n Y_n.$  This potential is 
then used 
in Bernoulli's law to determine the pressure in the 
liquid at the bubble
wall.  Applying the pressure jump condition across the 
interface yields the
Rayleigh-Plesset equation for $R(t)$ as well as a 
dynamical
equation for
the distortion amplitude $a_n(t)$, 
\begin{equation}
\ddot a_n + {3 \dot R\over R} \dot a_n -
\left[(n-1) \frac{\ddot{R}}{R} - {\beta_n \sigma \over 
\rho_w R^3} \right]
 a_n = 0,
\label{pro0}
\end{equation}
where 
$\beta_n = (n-1)(n+1)(n+2)$.
However, viscous effects have been neglected in 
Plesset's
derivation.

Viscosity was later taken into account by Prosperetti
\cite{pro77}. 
The intrinsic difficulty in its consideration is that
viscous stresses produce vorticity in the neighborhood 
of the bubble wall \cite{bat70}.
In principle, vorticity spreads both by convective 
and by diffusive
processes all over the fluid and the problem becomes 
nonlocal. However, for
small viscosity the generated vorticity will be more or 
less localized and we can introduce a bubble boundary 
layer
approximation of the nonlocal equations which we do in 
the next subsection.

Here, we give the dynamics of the nonlocal problem, 
closely following Prosperetti
\cite{pro77}.
It is advantageous to decompose the vorticity field in 
the fluid in a poloidal
and a toroidal part, which are conveniently represented 
by scalar fields
$S(r,t)$ and $T(r,t)$, respectively,
%\begin{eqnarray}
\begineq
\mb{\omega} = \mb{\nabla} \times \mb{\nabla} \times
\left[ S(r,t) Y_n^m (\theta , \phi ) {\bf e_r} \right] +
\mb{\nabla} \times \left[ T(r,t) Y_n^m(\theta , \phi ) 
{\bf e_r }\right] . 
\label{pro1}
%\end{eqnarrary}
\endeq
Only the latter, $T(r,t)$, contributes to the long term 
dynamics of the
bubble. Its dynamics is given by the PDE
\begineq
\partial_t T(r,t) + R^2 \dot R \partial_r \left( {1\over 
r^2} 
T\right) = \nu \partial_r^2 T - {\nu n (n+1) \over r^2 } 
T,
\label{pro2}
\endeq
by the nonlocal boundary condition
at $R(t)$,
\begineq
T(R, t) + 2 R^{n-1} \int_R^\infty s^{-n} T(s,t) ds =
{2\over n+1 } \left[ (n+2 ) \dot a_n - (n-1) {a_n\over R 
} \dot R \right],
\label{pro3}
\endeq
and by the boundary condition at infinity, $T(\infty , t 
) = 0$. At $t=0$
the fluid is assumed to be at rest.
The collapse $\dot R(t)$ of the bubble 
transports vorticity \mb{\omega} 
into the fluid, see eq.\ (\ref{pro2}).

Once created,
the vorticity acts back on the
dynamics of $a_n(t)$. These indirect viscous corrections 
together
with the direct ones modify eq.\ (\ref{pro0}) to yield 
\begin{eqnarray}
\ddot a_n &+& B_n(t) \dot a_n - A_n(t) a_n
+ n(n+1)(n+2) {\nu \over R^2} T(R,t)
\nonumber \\ &-& n(n+1) {\dot R \over R^2 } 
\int_R^\infty
\left[ 1- \left( {R\over s}\right)^3 \right] 
\left({R\over s}\right)^n
T(s,t) ds = 0
\label{pro4}
\end{eqnarray}
with
\begineq
A_n(t) = (n-1 ) {\ddot R \over R } - {\beta_n \sigma 
\over \rho_w R^3}
+ 2\beta_n \nu {\dot R \over R^3},
\label{pro5}
\endeq
\begineq
B_n(t) = {3\dot R \over R } - 2\beta_n {\nu \over R^2 }.
\label{pro6}
\endeq

\subsection{Boundary layer approximation}
For an exact stability analysis the coupled eqs.\ 
(\ref{pro2}) --
(\ref{pro4}) together with the RP equation (\ref{rp}) 
must be solved.
However, considerable vorticity is only to be expected 
in a small boundary
layer of thickness $\delta$ around the bubble.

Within this boundary layer approximation the space 
integrals in (\ref{pro3}) and
(\ref{pro4}) can be approximated by the (integrand at 
$R$)$\times\delta$.
The integral in (\ref{pro3}) thus is
$\approx R^{-n} T(R,t) \delta $ while the one in 
(\ref{pro4}) vanishes. We
obtain 
\begin{eqnarray}
\ddot a_n + B_n(t) \dot a_n - A_n(t) a_n=0
\label{pro8}
\end{eqnarray}
with
\begineq
A_n(t) = (n-1 ) {\ddot R \over R } - {\beta_n \sigma 
\over \rho_w R^3}
- {2 \nu \dot R \over R^3} \left[-\beta_n + n (n-1)(n+2) 
{1\over 1 + 2
{\delta \over R }}\right] ,
\label{pro9}
\endeq
\begineq
B_n(t) = {3\dot R \over R } +  {2\nu \over R^2 }
\left[ -\beta_n + {n (n+2 )^2 \over 1 + 2 {\delta \over 
R}}\right].
\label{pro10}
\endeq
The viscous contribution to
$A_n (t)$ is not important
and only causes a tiny shift,
as the ratio between the third and the second term of the rhs in
(\ref{pro9}) is typically $\nu\rho_w R_0\omega/\sigma \sim 10^{-2}$.
 However, in (\ref{pro10}) it introduces a 
damping rate
\begineq
\xi_n(t) = {2\nu\over R^2 } \left[ -\beta_n + {n 
(n+2)^2\over
 1+2 {\delta\over R}}  \right] ,
\label{pro11}
\endeq
acting on the shape oscillations.
That only the second term in (\ref{pro10}) contributes 
to the damping rate of the
oscillator can formally best be seen after the 
substitution \cite{ple54}
$b_n (t) \propto (R(t))^{3/2} a_n (t) $. 
Physically this is not surprising, as the first term is a mere
consequence of spherical geometry.
Two physical effects contribute to the damping rate: (i) 
Stabilizing, {\it local}
damping by viscous dissipation. In this limit
the viscous boundary layer around the bubble vanishes 
($\delta = 0$)
and the damping rate becomes $\xi_n (t) = {2\nu } (n+2) 
(2n+1)/R^2 >0$.
(ii) The
movement of eddies around the bubble,
generated by the shape oscillations itself. With 
increasing boundary layer
thickness $\delta$ (i.e., with increasing viscosity) 
this destabilizing
effect becomes stronger.

How to approximate the thickness $\delta$ of the 
boundary layer when we
have non-vanishing vorticity? 
For large bubbles $R\gg \delta $ it is set
by the diffusive length scale $\sqrt { \nu/\omega}=2.5\mu m$
in equation (\ref{pro2}) \cite{boundary}.
As typical frequency scale we choose the forcing 
frequency $\omega$.
Higher
frequencies are of course also present in the RP 
dynamics, but a Fourier
analysis of the R(t) signal shows that the forcing 
frequency is dominant.
Because of the angular contribution to the dissipation 
(the second term
on the rhs of eq.\ (\ref{pro2})) we also expect a 
slight dependence
on the spherical mode $n$ which we neglect here.

For small bubbles $R \ll \delta $ we do not
expect the boundary layer
around the bubble to be larger than the bubble itself. 
We thus have to introduce
a cutoff \cite{gro93}.
We choose
\begineq
\delta = min\left( \sqrt{{\nu \over \omega }}, {R\over 
2n} \right) .
\label{pro12}
\endeq
The n-dependence of the cutoff can be understood 
from
the quasi static limit which holds for small bubbles, as
for small bubbles the bubble dynamics is strongly damped
by viscosity and surface tension and $R(t)$ does not 
change much.
In this quasi static limit the lhs of (\ref{pro2}) 
vanishes and $T(r)
= T(R) (r/R)^{-n}$ is the
 static solution. It decays to half its boundary value
$T(R)$ at $r=2^{1/n}R$. Thus $\delta = R (2^{1/n} - 
1)\approx R \ln 2 /n
\approx R/(2n)$ as in (\ref{pro12}). More precisely,
for $T(r)= T(R)(r/R)^{-n}$ we can calculate $T(R)$ from 
eq.\ (\ref{pro3}),
and obtain essentially the same $\delta$.
The exact values of our results  depend on details of 
the cutoff
(\ref{pro12}).
However, the general features of the solution are 
invariant.

\begin{figure}[ht]
\setlength{\unitlength}{1.0cm}
\begin{center}
\begin{picture}(11,11)
%\put(0.5,7.5){\LARGE a)}
%\put(0.5,5){\LARGE b)}
\put(-.6,0.0){\psfig{figure=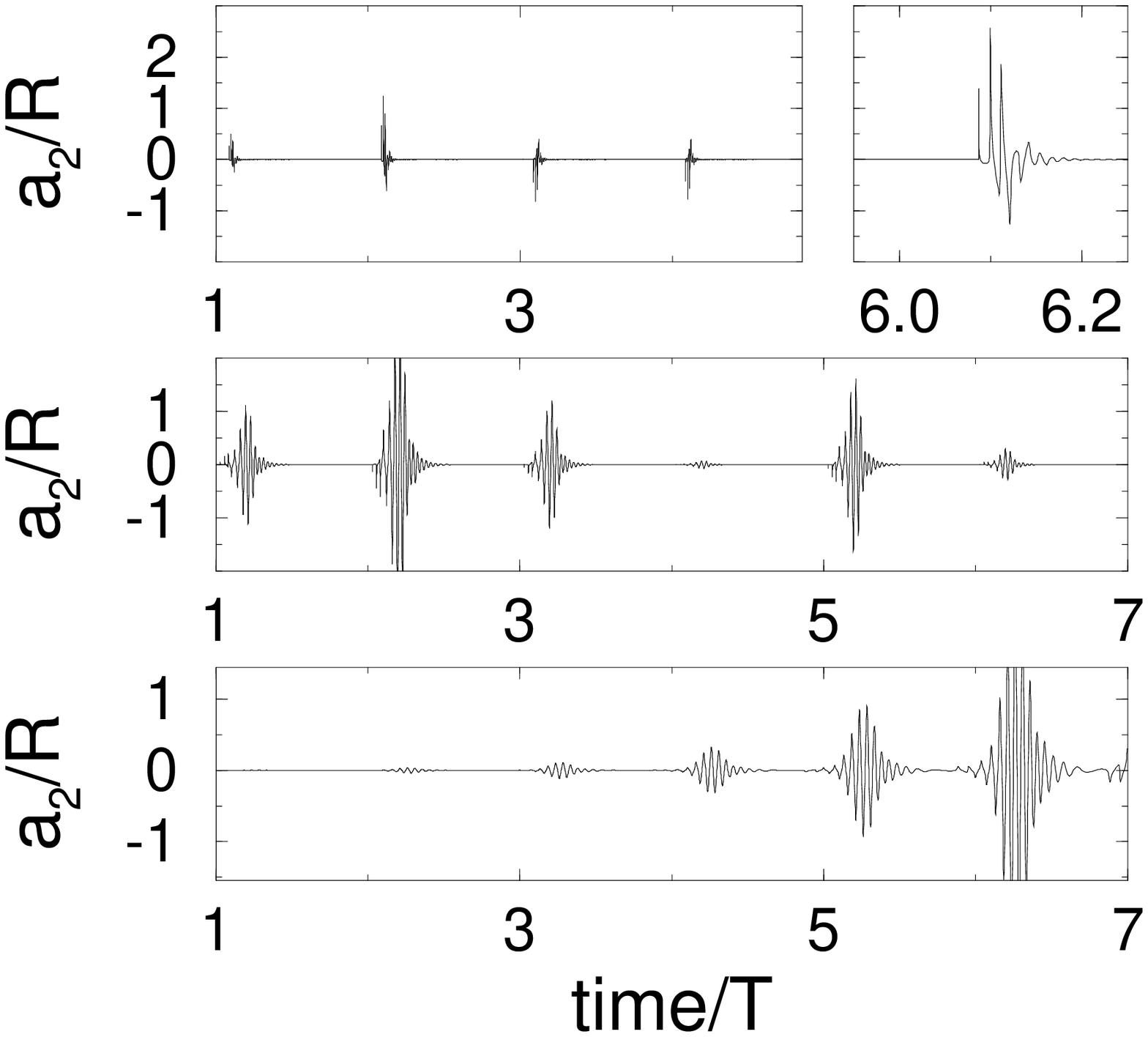,width=12cm,angle=0.}}
%\put(-0.6,0.0){\psfig{figure=dummy.eps,width=12cm}}
\end{picture}
\end{center}
\caption[]{
Time development
 of the normalized distortion amplitude $a_2(t)/R(t)$
for (a) 
a Rayleigh-Taylor unstable, parametrically stable bubble
($R_0 = 2.5\mu m$, $P_a=1.5$atm), (b)
an afterbounce unstable, parametrically stable bubble
($R_0 = 4.0\mu m$, $P_a=1.3$atm) and (c)
a Rayleigh-Taylor stable, parametrically unstable
bubble 
($R_0 = 5.2\mu m$, $P_a=1.0$atm). 
In (a) we also show $a_2/R$ in a blow up of the time scale
to demonstrate that the typical time scale of the Rayleigh-Taylor instability
is nanoseconds.
The typical time scales of afterbounce (b) and parametric instability (c)
is
in the  microsecond, and millisecond range, respectively.
}
\label{aoft}
\end{figure}

With the approximation (\ref{pro12}) we can 
Taylor-expand (\ref{pro9})
and (\ref{pro10}) and finally
obtain as approximate dynamical equation for
$a_n (t)$ equation (\ref{pro8}) with \cite{bre95}
\begineq
A_n(t) = (n-1 ) {\ddot R \over R } - {\beta_n \sigma 
\over \rho_w R^3}
- {2 \nu \dot R \over R^3} \left[
(n-1)(n+2) + 2n (n+2) (n-1)
{\delta \over R }
\right] ,
\label{pro13}
\endeq
\begineq
B_n(t) = {3\dot R \over R } +  {2\nu \over R^2 }
\left[
(n+2) (2n+1) - 2n (n+2)^2 
 {\delta \over R}
 \right].
\label{pro14}
\endeq
Our results are based on these equations.

\subsection{Rayleigh-Taylor, afterbounce, and parametric instabilities}
\noindent

Three types
of shape instabilities are hidden inside these equations. We call them
the Rayleigh-Taylor 
instability, the afterbounce instability,
 and parametric instability, for reasons which will become clear later.
They 
are distinguished by the widely different time scales over which they act.  
The transition between these instabilities is often gradual rather
than abrupt.
Nevertheless, we think that our classification is physically important as the 
difference in their time scales results in 
a difference of
the typical velocities of the bubble fragments after the shape instability
has destroyed the bubble.
We estimate this velocity as typical length scale $\sim 1\mu m$ of a
collapsed bubble divided
by the typical time scale of the pinch off.

The goal of this subsection is
to find criteria for the occurrence of the three shape instabilities.
Their nature
becomes clear from 
figure \ref{aoft}, where we show
the dynamics of $a_2(t)$ 
(which is the most unstable spherical mode in the parameter range
discussed in this paper, so we restrict ourselves to it)
for the three different regimes of instability.
We normalize $a_2(t)$ to the current bubble radius $R(t)$ to get
a measure of bubble distortion.
The upper part 
displays
the dynamics of the distortion amplitude $a_2(t)$,
normalized to $R(t)$, in
a Rayleigh-Taylor unstable,
parametrically stable case, obtained from a numerical solution of
eqs.\ (\ref{rp}), (\ref{pro8}), (\ref{pro13}), and (\ref{pro14}), with
addition of small amplitude noise.
The middle part shows the dynamics of the distortion
in the regime of the afterbounce
instability, and 
the lower part typifies the dynamics of a
parametrically unstable, Rayleigh-Taylor
stable bubble. Clearly,  very different
time scales are responsible for the distortion of the spherical shape
of the bubble. Of course
there are regimes in the $R_0 - P_a$ parameter space where the
bubble is stable (or unstable) towards two or even all three
instabilities.

First we focus on the Rayleigh-Taylor
 instability, occurring near the minimum bubble
radius when the gas accelerates into the fluid.  
The strongest destabilization occurs just after the bubble radius
reaches its minimum.
The acceleration of gas towards the fluid during this time is enormous, 
motivating the name Rayleigh-Taylor instability.
Closer analysis of this section of bubble movement \cite{bre96c} reveals 
that the time scale of the Rayleigh-Taylor
 instability can be estimated by the expansion
time scale of R(t) just after the collapse which is
$t_{RT}\sim {h/c_w}$ with the van der Waals hard core radius $h=R_0/8.86$.
Thus, $t_{RT}\sim 10^{-9} - 10^{-10}s$, which is confirmed by the numerical
results. 
In order to take into account microscopic fluctuations we added a random
displacement of size $\sim 0.1nm$ 
to the distortion $a_2(t)$ after each integration time step.

\begin{figure}[ht]
\setlength{\unitlength}{1.0cm}
\begin{center}
\begin{picture}(11,10)
%\put(0.5,7.5){\LARGE a)}
%\put(0.5,5){\LARGE b)}
\put(0.0,0.0){\psfig{figure=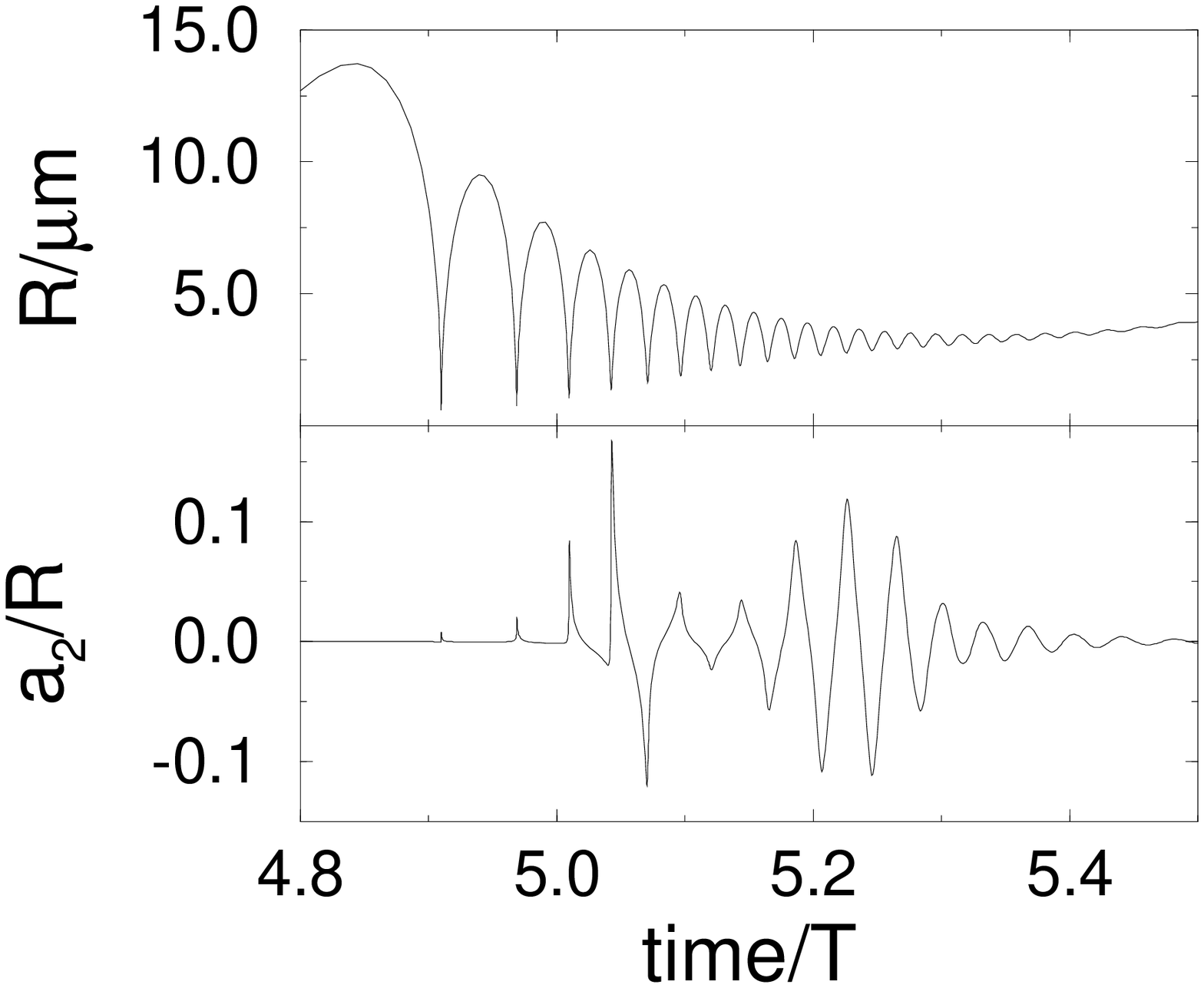,width=12cm,angle=-0.}}
%\put(0.5,0.5){\psfig{figure=dummy.eps,width=12cm}}
\end{picture}
\end{center}
\caption[]{
Time development
 of the bubble radius $R(t)$ (upper part) and
 distortion amplitude $a_2 (t)$ (lower part)
 for a $R_0=4.4\mu m$ bubble driven at $P_a=1.1$atm.
Note the transition from Rayleigh-Taylor (time scale
$ns$) to afterbounce perturbations (time scale $\mu s$) during the
afterbounce part of the bubble dynamics.
It is also seen that the dynamics of the distortion $a_2(t)$ has
half the frequency of its forcing bubble dynamics $R(t)$ as typical
for an instability of the Mathieu type.
}
\label{aoftrtab}
\end{figure}

For lower $P_a$ the destabilization during the violent bubble
collapse may not be strong enough to immediately overwhelm the bubble.
But as seen from figure \ref{aoft} further periods
of destabilization occur 
during the afterbounces. As pointed out above, in the afterbounce 
regime $R(t)$ oscillates on the bubble's intrinsic time scale $t_I \sim
\sqrt{\rho_w R_0^2 / P_0 } \sim 1\mu s$. This is too fast for viscous effects
to smooth out the shape distortions, so after a couple of afterbounces the
bubble may be overwhelmed. This type of instability shows features of
parametric instability, however, the afterbounces are not strictly periodic.
As an approximate
criterion for this {\it afterbounce instability} we give that 
microscopic fluctuations can overwhelm the bubble within one period T,
\begineq
\max_{\{t'| t<t'<t+T \}}
\left(
{|a_2(t')| \over  R(t')} 
\right) \ageq 1.
\label{rt_crit}
\endeq

The transition between Rayleigh-Taylor
 and afterbounce instabilities is illustrated
in figure \ref{aoftrtab}. The $a_2/R$ time series shows violent behavior
at the main bubble collapse and the first afterbounces. Then,
the behavior of $a_2(t)/R(t)$ becomes oscillatory and locks into the
periodicity of the $R(t)$ afterbounces with twice their period (the
same is true for $a_2(t)$ itself). This is
to be expected from a Mathieu type instability, which is most effective
for a driving with twice the intrinsic frequency of the driven equation.

Figure \ref{pi_rt} depicts a phase diagram of shape instabilities as a
function of the 
ambient bubble radius $R_0$ and the forcing pressure amplitude $P_a$ for
$n=2$. The dashed line gives the combined stability threshold for
Rayleigh-Taylor and
afterbounce instabilities (the ``fast'' instabilities with time scales
$\ll T$).
The global features of the phase diagram are easily understood. Small
bubbles are more stable than large ones thanks to surface tension and
viscosity, as the second term in (\ref{pro14}) becomes dominant
for small radii $R$. Evidently, weakly forced bubbles are more stable
than strongly forced ones.

A {\it pure}
parametric shape instability
acts on the much longer time scale of the forcing $T\approx 38\mu s$. 
It corresponds to
a net growth of a nonspherical perturbation
over one oscillation period, so that after many
periods perturbations overwhelm the bubble. 
The time scale of the parametric instability $t_{PI}$ is thus many
forcing periods $T=2\pi/\omega$.
However, as the finally resulting pinchoff occurs during afterbounces, its
time scale is
the same as for the afterbounce instability, i.e., the intrinsic
time scale of the bubble motion $t_I \sim 1\mu s$.

In the relevant parameter regime 
for the parametric instability $R(t)$ and thus also
$A_n(t)$ and $B_n(t)$ are strictly periodic in time with frequency
$1/T$.
Thus 
Eq.\ (\ref{pro8}) is  an ODE of Hill's type
and the 
parametric instability 
can be rigorously analyzed. It 
occurs whenever
the magnitude of the maximal eigenvalue of
the Floquet transition matrix $F_n(T)$ of eq.\ (\ref{pro8})
is larger than one. 
The Floquet transition matrix
$F_n(T)$ is defined by 
\begin{equation}
\left(\begin{array}{c}
a_n(T) \\ \dot a_n(T)
\end{array}\right)= F_n(T)
\left(\begin{array}{c}
a_n(0) \\ \dot a_n(0)
\end{array}\right).
\end{equation}
By numerically computing the eigenvalues of the Floquet transition matrix we
mapped out the phase diagram of stability.  Fig.\ \ref{pi_rt}
shows the stable and unstable domains in the $R_0 - P_a$ parameter space.

In the SL parameter range of $P_a \approx  1.2$ to $1.5$atm
the bubble becomes 
parametrically unstable at about $R_0^{PI} \approx 4 - 5 \mu m$. This number
is not to be understood as a prediction of the exact value, as within our
approximations we can only predict the order of magnitude and trends. 

For smaller $P_a$ the threshold for instability $R_0^{PI}$ does depend
on the forcing pressure. We discussed phase diagrams in those regimes
in \cite{bre95} and
 also showed that in the small forcing limit
 eq.\ ({\ref{pro8}) reduces 
to a Mathieu equation.

\begin{figure}[ht]
\setlength{\unitlength}{1.0cm}
\begin{center}
\begin{picture}(11,10)
%\put(0.5,7.5){\LARGE a)}
%\put(0.5,5){\LARGE b)}
\put(0.0,0.0){\psfig{figure=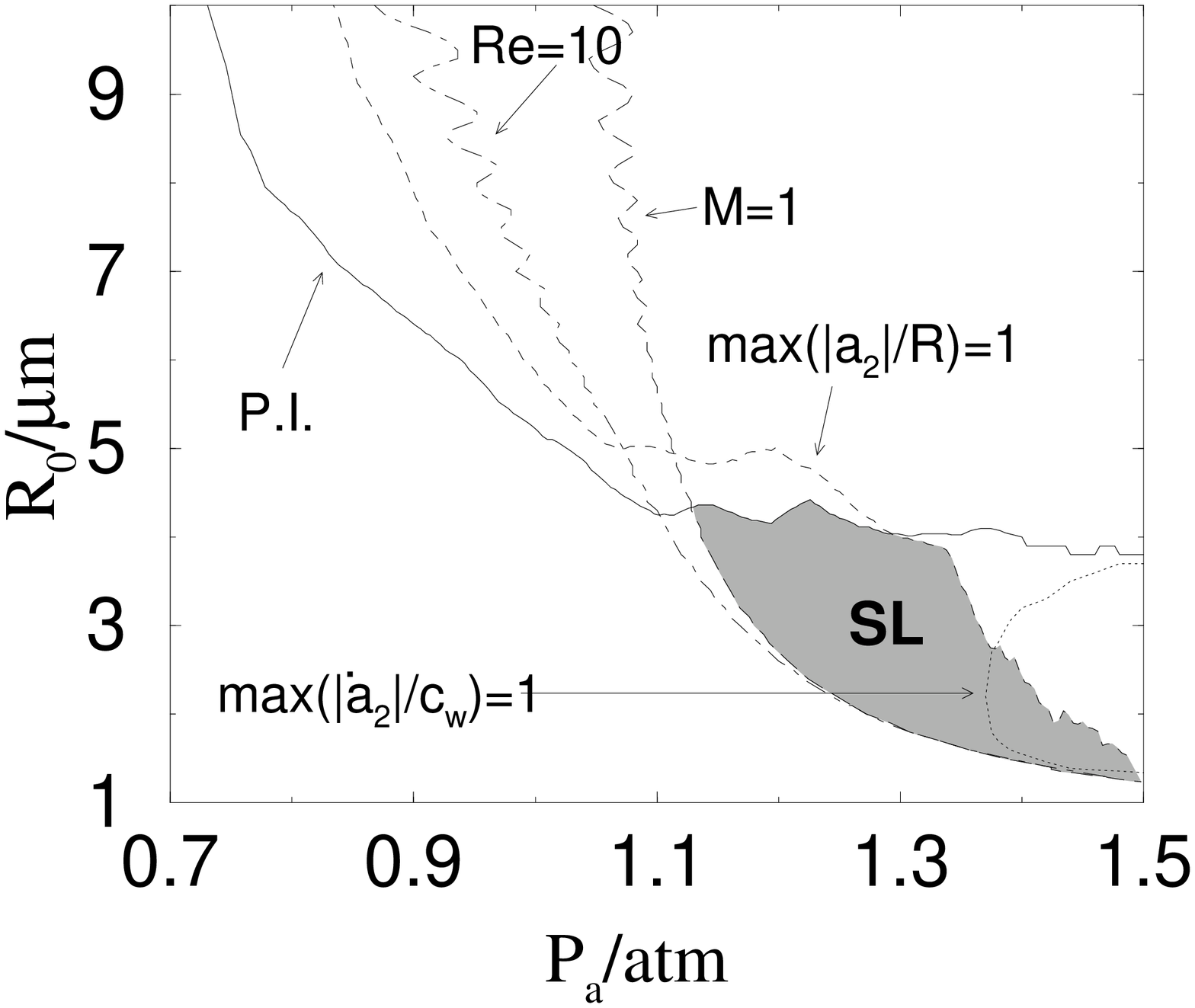,width=12cm,angle=-90.}}
%\put(0.0,0.0){\psfig{figure=dummy.eps,width=12cm}}
\end{picture}
\end{center}
\caption[]{
Borderline of the parametric instability (solid), the
afterbounce instability according to criterion
(\ref{rt_crit}) (short dashed), and 
the $M=1$ criterion (\ref{mach}) for a supersonic bubble collapse
(long dashed). The $Re=10$-
criterion (\ref{re}) for the persistence of a shock 
(dot-dashed) is found to be less stringent than the $M=1$ criterion.
Also shown is the
perturbation velocity threshold
(\ref{adot_crit}) (dotted). In case of unstable SL, 
to the right of this curve the bubble is thrown out of the trap as the
pinch off of micro bubbles is too violent. The region where SL is
possible is shaded. 
}
\label{pi_rt}
\end{figure}

That $R_0^{PI}$ does not significantly depend on $P_a$ for large $P_a$ can be
understood from the dynamics of $R(t)$ and $a_n(t)$ and from equation
(\ref{pro14}). If $R(t)$ is small, the second term in (\ref{pro14})
dominates and stabilizes $a_n(t)$. For small $P_a$ the minimal radius
$R_{min} = min_t (R(t))$ still decreases with increasing $P_a$. But after
the van der Waals hard core radius $h=R_0/8.86$ has once been reached
for large enough $P_a$,
$R_{min}$ becomes $P_a$ independent
\cite{bre96c}.

All calculation have been performed for the viscosity of water
$\nu = 10^{-6}m^2/s$.
Of course
$R_0^{PI}$ and the other thresholds strongly depend on
$\nu$. E.g., for a viscosity five times
that of water we have $R_0^{PI} \approx 10 \mu m$,  but we won't discuss
this dependence here.

\begin{figure}[htb]
\setlength{\unitlength}{1.0cm}
\begin{center}
\begin{picture}(12,8)
%\put(0.5,7.5){\LARGE a)}
%\put(0.5,5){\LARGE b)}
%\put(0.0,-5.0){\psfig{figure=figure6.ps,width=14cm,angle=0.}}
\put(0.0,-5.0){\psfig{figure=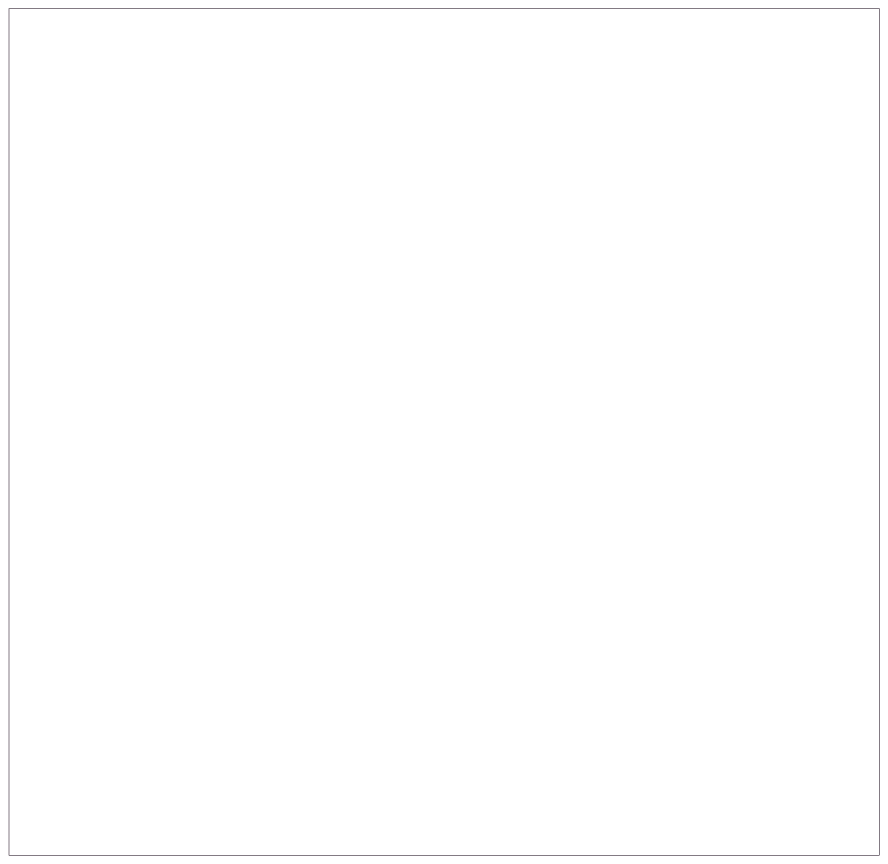,width=14cm}}
\end{picture}
\end{center}
\caption[]{
Maximum perturbation velocity according to (\ref{adot_crit}) as a
function of $P_a$ and $R_0$. This calculation was done with an $a_2$ noise
amplitude of $1nm$. Note the two distinct regions of the
parametric instability
(left) with its rapid onset 
and the Rayleigh-Taylor instability (right, high pressures and small $R_0$).
In the latter region the bubble is thrown out of the trap
when the pinch off occurs. 
It corresponds
to the region right of the dotted line in figure \ref{pi_rt}.
}
\label{3dadot}
\end{figure}

\subsection{After the shape instabilities}
All types of shape instabilities result in the pinching off of
micro bubbles.  In other experimental configurations such
as the Faraday experiment \cite{ll87}, parametric
instability can saturate at a finite amplitude.
This is also possible for larger bubbles which are driven with
small forcing pressure $P_a < 1$atm \cite{gai90,bre95b}.
However, we believe the nonlinear saturation
is unlikely in the present experiment because the bubble size changes
by two orders of magnitude during a single oscillation period.   
Saturation would require that the amplitude $a_n$ is much smaller than
the {\it minimum} radius $R_{min}$; however, since the bubble spends 
most
of the cycle with $R \gg R_{min}$, the nonlinearities mainly act at larger
radii.

The major question, therefore, is what happens after the pinch off of
a micro bubble.  Will the bubble remain trapped at the anti--node
of the pressure field, or will it escape from the system?

The force holding the bubble in the trap is
the so-called Bjerknes force given by \cite{bre95b,bla49b}
\begineq {\bf F} = -  {4\over 3} \pi R^3 \mb{\nabla}P  .
\label{force}
\endeq
Both $R(t)$ and $\mb{\nabla}P(t)$
are oscillating with
time. The combination of both will lead to an effective, period
averaged force $\left< {\bf F}\right> $ which pushes bubbles smaller than
the resonance radius ($\sim 100\mu m$ here)
to the pressure antinode \cite{bre95b}. 
The bubbles are thus trapped.

During the collapse the forcing pressure is positive and the force
(\ref{force}) repulsive. If micro bubbles pinch off at this instant,
will they and
the recoiled bubble be fast enough to escape from the node before the
force becomes attractive again? If so, they must travel a quarter of
the acoustic wavelength $c_wT/4$ in time $T/2$, thus their velocity must
be of order $\sim c_w\sim 10^3m/s$.
Assuming a typical length scale of $R_0\sim 1\mu m$ for the micro bubble
and the remainder, we obtain as critical time scale of the collapse
$10^{-9}s$: If the collapse is faster, the bubble can not survive in the
trap, if it is slower, it is likely to survive as a smaller bubble.
From the estimates
of the relevant collapse time scale in the last subsection we conclude that
the bubble will survive the afterbounce and the parametric instability
where the pinch off occurs on the intrinsic time scale
$t_I \sim 1\mu s$.
The pinched off micro bubble from those shape instabilities will
dissolve by diffusion (see below), but the remainder of the bubble can survive
and grow by rectified diffusion. 

The bubble fragments can, however, be ejected from the trap after a
RT instability ($t_{RT}\sim 10^{-9}s$). 
To get a more stringent criterion for ejection, we computed the maximal
velocity $| \dot R + \dot{a}_2|$. For a bubble split apart by
shape instabilities, this will also give the typical fragment velocity.
The outward velocity is dominated by the $\dot a_2$ term;
$\dot R$ does not exceed $0.1 c_w $ \cite{bre96c}.
Therefore, we assume 
that the fragments will escape if
\begineq
\max_{\{t'| t<t'<t+T \}}
\left(
{|\dot{a}_2(t')| \over c_w} 
\right) \ageq 1\,.
\label{adot_crit}
\endeq

Figure \ref{3dadot} depicts the lhs of this inequality as a function of
$P_a$ and $R_0$. The perturbation speed exceeds the sound speed in two
clearly distinct 
regions of parameter space: at large radii, where the parametric
instability leads to high velocity, and at small radii and high pressures,
where $c_w$ is reached during Rayleigh-Taylor
 instability. Of course, only the latter
region will determine the escape of the bubble, because parametrically
unstable
bubbles shed micro bubbles long before these high velocities are reached.
We extract a $|\dot{a}_2| = c_w$ isoline from this graph and add it to
the phase diagram in figure \ref{pi_rt} (dotted line). In accord with
the estimate for $t_{RT}$ in the preceding section, bubbles with
smaller $R_0$ have faster instability time scales and thus higher values
of $|\dot{a}_2|$. However,
for very small $R_0$, the bubble dynamics is stabilized
by surface tension, and the perturbation velocity drops again. The exact
position of this line is not meant to be quantitative, because it is
extracted from dynamical data for bubble velocities where the RP equation
is not a good approximation to bubble dynamics any more.

\section{Energy Focusing Mechanism}\label{shocks}
\setcounter{equation}{0}

Although this paper is primarily concerned with stability constraints
on a bubble obeying the Rayleigh Plesset equation, in order to relate the
calculations to the sonoluminescence experiments it is necessary to
adopt an onset criterion for the light emission.  The light production
is triggered by a {\it hydrodynamic} energy focusing mechanism, which
concentrates the input energy enough so that light is
produced.  The mechanism through which the focused energy produces light
is outside the scope of this paper;
many suggestions such as bremsstrahlung or black body
radiation are present in literature \cite{wu93,loe93}. 

Two theories of hydrodynamic energy focusing have been proposed:  The
original theory was that
during the collapse of the bubble shocks detach
from the gas-water interface and focus to the bubble's center
\cite{gre93,wu93}.  The motion of a focusing shock is described by
Guderley's similarity
solution to the hydrodynamic equations \cite{gud42,ll87}, which dictates that
the temperature at the shock diverges as $T\propto R_s^{-p}$, where
$R_s$ is the distance of the shock from the origin and $p\sim 1$
is an irrational
scaling exponent.   The amount of energy focusing in this theory is
determined by the minimum distance $R_{s,min}$ to which the shock
approaches the origin.

The second theory of energy focusing
\cite{bre96d} posits that the bubble is an ``acoustic resonator'',
and that acoustic energy builds up in the bubble over
many oscillation periods.  Within this picture
the amount of energy focusing is set by the total stored
energy in the bubble.

The crucial issue for the present paper is the onset criterion for how
strongly the bubble must be forced for significant energy focusing
to occur.
It turns out that
both shocks and collective energy buildup require the
same onset criterion:  the inward bubble wall velocity must be of
order of the speed of sound in the gas,
\begineq
M= {-\dot R\over c_{gas} } \ageq 1.
\label{mach}
\endeq
Here, $M$ is the Mach number, and
\begineq
c_{gas}^2
= \gamma {p\over \rho}
{R^3 \over R^3-h^3}
\label{cgas}
\endeq
is the (squared) velocity of sound
in a van der Waals gas.
For the shock wave theory the criterion (\ref{mach}) is motivated
by the fact that shocks can only detach from the bubble wall
if its velocity is supersonic
\cite{ll87,loe93}.
For the acoustic resonator theory
\cite{bre96d} the criterion (\ref{mach}) holds as resonators
can only be supplied with energy if they are forced on a time scale
comparable to the intrinsic time scale. Here, the forcing time scale
is $R/|\dot R|$ and the intrinsic acoustic
resonator time scale is $R/c_{gas}$.
Thus $|\dot R |/c_{gas} \sim 1$. From above discussion of the time scales
in the RP equation it follows that all time scales between $R/|\dot R|$
and $T$ are present in the $R(t)$ dynamics; thus (\ref{mach}) is
the correct criterion for energy focusing.

For large enough forcing 
the energy focusing
 criterion (\ref{mach})
 is fulfilled once per cycle, namely shortly before the bubble achieves
its minimum radius which in the relevant $P_a$ domain is very close to
the hard core radius $h$. 
This is where the light pulse is emitted \cite{gai90,bar91,cru94}
which gives support to the criterion (\ref{mach}). In figure
\ref{pi_rt} we plotted the threshold
for the 
$M=1$ criterion in the $P_a-R_0$ parameter domain.

Let  us check two further conditions which should be fulfilled 
within the shock wave theory. First, another 
requirement besides (\ref{mach})
is that the shock must persist. Dissipative mechanisms
inside the bubble must therefore be weak.
A measure for the relative strength of inertial and dissipative
mechanisms is the Reynolds number $Re$. Viscous effects
dominate for small $Re$. As crossover for nonlinear effects such
as shocks to take
over we take \cite{ll87}
\begineq
Re = {R |\dot R |\over \nu_{gas}} \ageq 10.
\label{re}
\endeq
The kinematic viscosity of argon is $\nu_{gas} = 11 \cdot 10^{-6}m^2/s$.
We neglect its temperature dependence. In the center of the bubble
this is a poor approximation and $\nu_{gas}$ will be lower.
The $Re>10$ criterion will then be fulfilled earlier. However, 
(\ref{mach}) is the more stringent criterion anyhow as seen
from figure \ref{pi_rt}.

Second, we have to compare the thickness of a shock with the bubble's size.
The thickness of a shock is of the same order of magnitude
as the mean free path $l$ of a gas molecule \cite{ll87}.
We have $l\sim V/(N\sigma_0)$ where $V$ is the bubble volume,
$N$ the number of particles in the bubble, and $\sigma_0 \sim 10^{-19}
m^2$ the collision cross section of argon atoms.
Thus
\begineq
l=l_0 \left( {R\over R_0 }\right)^3 .
\label{l}
\endeq
With $l_0 \sim 10^{-7} m$ we obtain
$l\sim 10^{-10}m$ during the strongest compression, i.e., a very sharply
defined shock with a width $\ll R$ in spite of viscosity.

We now come back to figure \ref{pi_rt}. That plot summarizes
the criteria we suggest to be necessary for SL to occur:
(i) Bubble wall Mach number $M>1$ to ensure energy focusing 
to reach the high temperatures necessary for SL.
(ii) Short time scale shape stability (Rayleigh-Tarylor
 and afterbounce) and 
(iii) parametric stability.
(iv) Finally, the perturbation speed must not exceed $c_w$, in order
to keep the bubble or its fragments trapped in the sound field.

There is only a small domain in parameter space where
the bubble fulfills all four criteria. 
This domain is shaded in fig.\ \ref{pi_rt}.
It is this domain where we expect SL to be possible (within our
RP-SL approach).
However, up to now no statement on the {\it diffusive stability}
has been made. We will address this subject in the next sections
and find that for {\it low enough gas concentration} the bubble in the shaded
domain is also diffusively stable.

\section{Diffusive Stability}\label{diffstab}
\setcounter{equation}{0}

Two types of SL in argon bubbles have been observed \cite{bar95}: For large
argon concentrations 
$P_\infty = 200$mmHg or 50mmHg the SL bubble is
diffusively unstable whereas for low concentrations $P_\infty = 3$mmHg the
bubble is diffusively stable and the relative phase of light emission stays
constant for days, see figure \ref{phi_fig}c
for the experimental result. 
With the ambient pressure $P_0=760$mmHg = 1atm these three gas concentrations
translate into relative concentrations of $c_\infty/c_0 = 0.26$, 0.06, and
0.004, respectively, where $c_0=0.061 kg/m^3$
is the saturation (mass) concentration of argon in
water for room temperature.

In this section we set out to quantitatively understand the difference
between the high and low concentration. Our
goal is to calculate a {\it phase diagram} in the parameter space of the two
experimental control parameters concentration $c_\infty / c_0$ and forcing
$P_a/P_0$. For given concentration and given forcing we will thus be able to
predict which of the three phases ``diffusively stable SL'', ``diffusively
unstable SL'', and ``no SL'' will be realized. The third parameter, the ambient
bubble radius $R_0$ is not at the experimenter's disposal but the system will
choose $R_0$ itself.
The ambient radius will follow from our analysis. First, we
will present phase diagrams in the ambient radius -- concentration and
ambient radius -- forcing pressure phase space.

We again stress that it is the application of the {\it adiabatic 
 approximation} of the diffusive problem \cite{fyr94,loe95}
which allows for the exploration of the whole 3D phase space $R_0 - c_\infty -
P_a$. Strictly speaking,
it only holds in the limit of zero diffusion constant
$D\to 0$. However, in the next section we will show that for the physical
diffusion constant $D= D_{Ar} = 2\cdot 10^{-9} m^2/s$ the deviations
between the exact solution and the adiabatic approximation are tiny, so
that we can apply it here.

\subsection{Formulation of the diffusive problem}
Assuming spherical symmetry, the mass concentration of gas
$c(r,t)$ dissolved
in the liquid at distance $r>R(t) $ from the center of the bubble
obeys the 
advection diffusion equation
\begin{equation}
r^2\partial_tc + R^2 \dot{R} \partial_r c = D\partial_r( r^2\partial_r c).
\label{convdiff}
\end{equation}
As boundary condition at the bubble wall we assume  
Henry's law
\begineq
c(R,t)= c_0 p(R,t)/P_0.
\label{henry}
\endeq
The concentration at $r\to\infty$ is given by 
$c_\infty$,
\begineq
c(\infty , t) = c_\infty.
\label{bc2}
\endeq
The concentration gradient at
the moving boundary gives the mass loss/gain of the bubble
\begineq
\dot{m} =  4 \pi R^2  D \partial_r c|_{R(t)}.
\label{gainloss}
\endeq
The bubble is driven by the RP equation (\ref{rp}). Together with the
initial conditions $R(t=0)=R_0$, $\dot R(t=0) = 0$ and $c(r,t=0)=c_\infty$ this
set of equations completely defines the problem.

With the  transformation
to the Lagrangian coordinate \cite{ell69}
\begineq
h(r,t) = {1\over 3} \left( r^3 - R^3(t)\right)
\label{trafo1}
\endeq
the advection diffusion PDE (\ref{convdiff}) simplifies to the 
diffusive equation
\begineq
\partial_t c(h,t) = D \partial_h \left[\left(3h+R^3(t)\right)^{4/3}
\partial_h c(h,t)\right].
\label{pde2}
\endeq
Eq.\ (\ref{pde2}) can still not be solved analytically. How it can best
be solved numerically is sketched in appendix \ref{appa}.

\subsection{Adiabatic approximation}
The main idea of Fyrillas and Szeri \cite{fyr94}
and L\"ofstedt et al.\ \cite{loe95}
is to treat the diffusive PDE by the method
of separation of time scales \cite{hin91}.
They split the concentration field in an
oscillatory part $c_{osc}(r,t) $ changing on the (fast) time scale $T$ of the
driving field and a smooth part $c_{smo}(r,t)$ changing on  a slow diffusive
time scale $\tau_D \gg T$,
\begineq
c(r,t) = c_{osc}(r,t) + c_{smo}(r,t).
\label{split}
\endeq
This approach can be thought of as
having introduced an ``adiabatic'' or slow time $\bar t$. The smooth
profile $c_{smo}$ only depends on the adiabatic time, $c_{smo}(r,\bar t)$. In
the PDE for $c_{smo}(r,\bar t)$ the fast time scale $\sim T$ is averaged out.
We define
$\tau_D=R_0^2/D$ as diffusive time scale.
Then the
Schmidt number $Sc= 2\pi \tau_D/T$ is a measure of
the quality of time scales separation. $Sc\to \infty$ or $D\to 0$ means
perfect separation.

It turns out to be useful to introduce {\it weighted} time averages,
\begineq
\left<f(t)\right>_{t,i} =
{\int_0^T f(t) R^i(t) dt \over
\int_0^T R^i(t) dt},
\label{av}
\endeq
which may still depend on the adiabatic time $\bar t$. Here, one only needs
$\left< . \right>_{t,0}$ and 
$\left< . \right>_{t,4}$.

The main result of ref.\ \cite{fyr94} is that in the asymptotic limit
$\bar t \to \infty$ the smooth profile $c_{smo}(h,\bar t)$ converges to
\begineq
\bar c_{smo} (h)  =  c_\infty + 
\left[ c_0 {\left< p(t)\right>_{t,4} \over P_0}
-c_\infty\right] \cdot   
\left\{1-
{
\int_0^h {dh'\over \left<(3h'+R^3(t))^{4/3}\right>_{t,0}}
\over
\int_0^\infty {dh'\over \left<(3h'+R^3(t))^{4/3}\right>_{t,0}}
}
\right\}. 
\label{fsmain}
\endeq
{}From (\ref{fsmain}) the adiabatic growth of the bubble can be calculated
as
\begineq
{d\over d\bar t} R_0(\bar t) =
{D c_0\over
\rho_0 R_0^2 (\bar t )} 
{
\left[ {c_\infty \over c_0} - {\left< p(t)\right>_{t,4} (\bar t)\over P_0}
\right]
\over
\int_0^\infty {dh'\over \left<(3h'+R^3(t))^{4/3}\right>_{t,0}} 
} .
\label{fsgrowth}
\endeq
The determination of the adiabatic growth rate has thus been reduced to solving
the RP ODE (\ref{rp}) for $R(t)$, calculating time averages $\left< .
\right>_{t,i}$ of functions of $R(t)$, and the solution of an space
integral. We
thus understand the adiabatic
approximation as being in the spirit of the RP approach.

\begin{figure}[htb]
\setlength{\unitlength}{1.0cm}
\begin{center}
\begin{picture}(11,9)
%\put(0.5,7.5){\LARGE a)}
%\put(0.5,5){\LARGE b)}
\put(-1.0,0.0){\psfig{figure=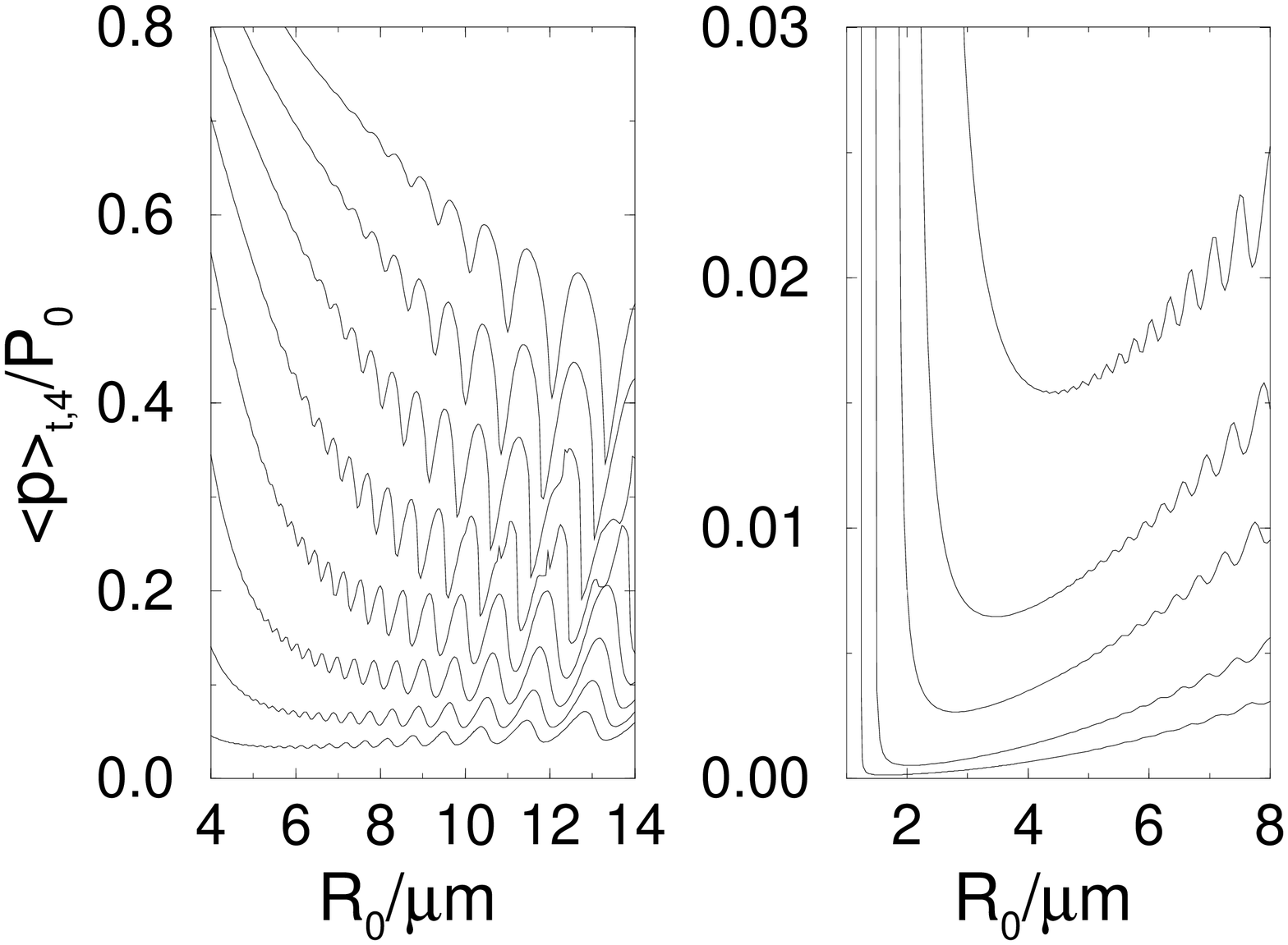,width=12cm,angle=-90.}}
%\put(-1.,0.0){\psfig{figure=dummy.eps,width=12cm}}
\end{picture}
\end{center}
\caption[]{ 
$\langle p\rangle_{t,4}/P_0$ as a function of
the ambient radius $R_0$ for small
forcing pressure amplitudes $P_a$=0.8atm to $P_a$=1.15atm (left, top to bottom, in
steps of 0.05 atm)
and for large $P_a$=1.2, 1.25, 1.3, 1.4, 1.5atm (right, top to bottom).
Note the different ordinate scales.
}
\label{p_tau}
\end{figure}

\subsection{Equilibrium points}
We now apply eq.\ (\ref{fsgrowth}) to the acoustically driven argon bubble
\cite{bar95}.
Let us disregard the shape instabilities discussed
in section \ref{shapestab} for the time being. 

The ambient bubble radius is in equilibrium (within the adiabatic
approximation), if
\begineq
{c_\infty\over c_0} = {\left< p(t) \right>_{t,4}\over P_0}.
\label{equi}
\endeq
Note that
within the adiabatic approximation
the condition does not depend on the diffusion coefficient $D$.
However, the smaller $D$,  the better the adiabatic approximation holds.
The equilibrium is stable, if
\begineq
\beta =  {d\left< p(t) \right>_{t,4}\over dR_0}
\label{stab}
\endeq
is positive.
Scaling laws for
$\left< p\right>_{t,4}$
and $\beta$ are discussed in
the next subsection and in ref.\ \cite{bre96c};
here we concentrate on the physical consequences.

\begin{figure}[htb]
\setlength{\unitlength}{1.0cm}
\begin{center}
\begin{picture}(11,10)
%\put(0.5,7.5){\LARGE a)}
%\put(0.5,5){\LARGE b)}
%\put(0.0,0.0){\psfig{figure=figure8.eps,width=12cm,angle=-90.}}
\put(0.0,0.){\psfig{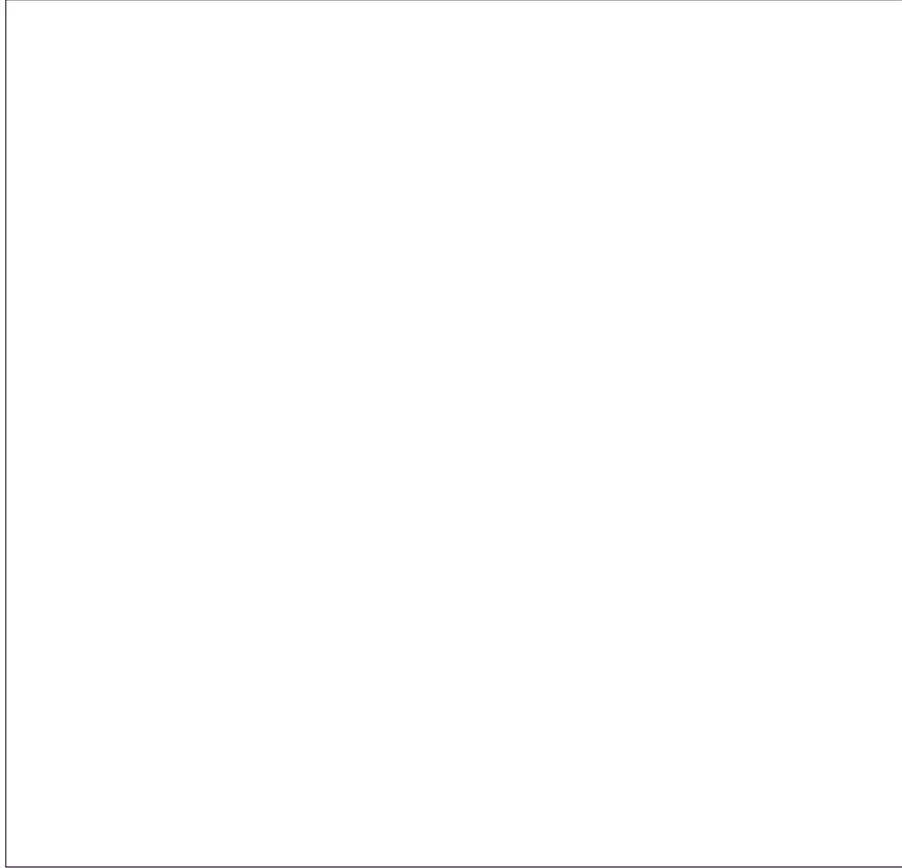}}
\end{picture}
\end{center}
\caption[]{ 
Bifurcation diagrams in the $R_0 - c_\infty$ parameter plane
for a forcing pressure of $P_a=1.3$atm (left) and  $P_a = 1.1$atm (right).
Tangent bifurcations are seen. The regimes with positive slope are
stable.
To the left of the curves the bubbles shrink,
to the right of them they grow.
}
\label{bif_dia_c}
\end{figure}

In figure \ref{p_tau} we plot
$\left< p (t) \right>_{t,4}$ as a function
of $R_0$ for various forcing pressure amplitudes $P_a$.
We first focus on small
$P_a\approx 0.8$atm and high gas concentration of $c_\infty/c_0\approx 0.7$.
There exists an {\it unstable} equilibrium at $R_0 \approx 6\mu m$.
Smaller bubbles shrink and finally dissolve, larger bubbles grow by rectified
diffusion.
For larger $P_a$ the average
$\la p(t)\ra_{t,4} (R_0)$ behaves quite differently 
in the small $R_0$ regime. It starts to show
characteristic wiggles, which can also be seen in $R_{max} (R_0)$.
Here, $R_{max}$ is the maximal radius over one period,
$R_{\rm max}( t) = {\rm max} \{R(t')| t\le t'\le t+T \}$.
The origin of the wiggles is a kind of resonance phenomenon in
the RP equation and can quantitatively
be understood in detail \cite{bre96c}.
Here we only discuss their consequences for the oscillating bubble.
They mean that the bubble may stabilize through a tangent bifurcation:
Imagine a fixed forcing $P_a$ and then decrease $c_\infty/c_0$. The
$c_\infty/c_0$ line will finally touch $\la p\ra_{t,4}/P_0$ at  
a local maximum and create a pair of stable and unstable fixed points
which will separate
for decreasing $c_\infty/c_0$. This process can repeat many times. The
fixed points
vanish through inverse tangent bifurcations, i.e., stable and unstable fixed
points
merge. The full bifurcation diagram is shown in figure \ref{bif_dia_c}
for $P_a=1.1$atm and for $P_a=1.3$atm.

The interpretation of the
$R_0$ - $c_\infty$  
phase diagram is as follows. The line signals
equilibrium for the
given driving pressure $P_a$.
It is stable if the slope $\partial
R^e_0/\partial c_\infty
|_{P_a}$ is positive, and unstable if it is negative. To the left of the line
the bubbles shrink, to the
right of it they grow by rectified diffusion. The shrinking
or growing bubbles can hit a stable fixed point and thus stabilize. The basin
of attraction of the stable fixed points is considerably larger for larger
forcing $P_a$, see figure \ref{bif_dia_c}.

\begin{figure}[htb]
\setlength{\unitlength}{1.0cm}
\begin{center}
\begin{picture}(11,9.3)
%\put(0.5,7.5){\LARGE a)}
%\put(0.5,5){\LARGE b)}
\put(0.0,0.0){\psfig{figure=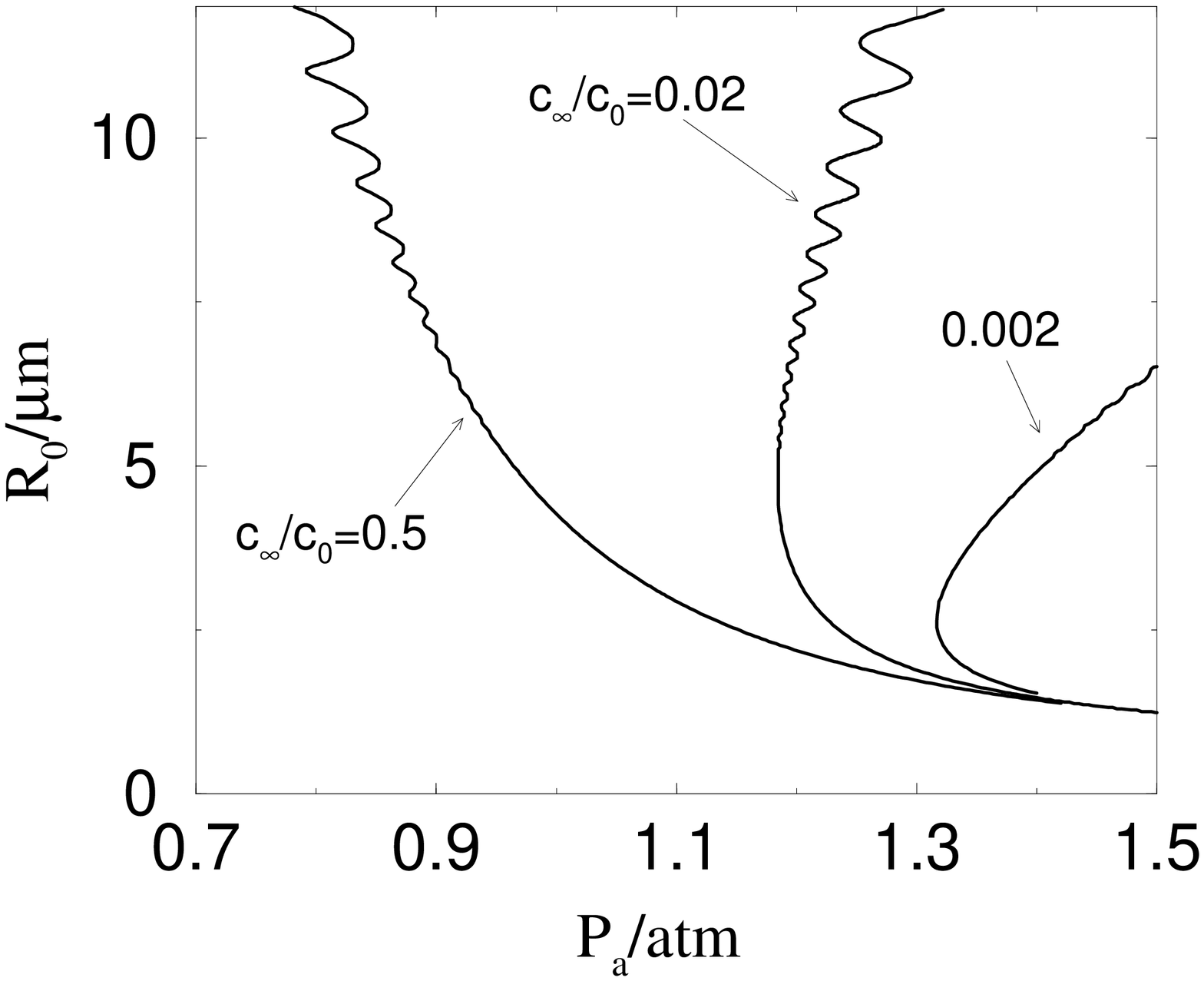,width=12cm,angle=-90.}}
%\put(0.5,0.5){\psfig{figure=dummy.eps,width=12cm}}
\end{picture}
\end{center}
\caption[]{ 
Bifurcation diagrams in the $R_0 - P_a$ parameter space.
Tangent bifurcations are seen. The regimes with positive slope are
stable. Gas concentrations are $c_\infty/c_0=0.002$ (right),
$c_\infty/c_0=0.02$ (middle), and   
$c_\infty/c_0=0.5$ (left).  
To the left of the curves the bubbles shrink and finally dissolve,
to the right of them they grow by rectified diffusion.
}
\label{bif_dia}
\end{figure}

A similar looking phase diagram results
when fixing $c_\infty/c_0$ and varying $P_a$.
In figure \ref{bif_dia} we show such a bifurcation diagram in the 
$R_0-P_a$ parameter space
for large concentration
$c_\infty/c_0=0.5$, for lower concentration
$c_\infty/c_0=0.02$, and
for very low concentration
$c_\infty/c_0=0.002$.
Again, the lines signal equilibrium; stable equilibrium for positive slope
$\partial
R^e_0 / \partial P_a |_{c_\infty}$, unstable equilibrium for negative
slope. To the
right of the line we have growth, left of it shrinking.
Increasing $P_a$ at fixed $c_\infty$ again leads to stabilization through a
series of tangent bifurcations and later on destabilization by inverse
tangent bifurcations.

From these figures \ref{bif_dia_c} and \ref{bif_dia} 
we immediately understand why there is
{\it no} diffusively stable SL for {\it large} Ar
concentration \cite{bre96}. No
diffusively stable bubble radii exist in the large $P_a$ - small
$R_0$ parameter regime where the
energy focusing condition for SL (\ref{mach}) is fulfilled.
Note that there are stable
equilibria, but for smaller $P_a$ and large $R_0$. Note also that the basin of
attraction of these stable equilibria is tiny.

For small concentration $c_\infty / c_0$ the situation is quite different. As
seen from figure \ref{bif_dia_c} and \ref{bif_dia} now there are
stable equilibria in the high $P_a$ -- low $R_0$ regime where the bubble is
both stable towards shape oscillations and fulfills the energy focusing
criterion.

\begin{figure}[htb]
\setlength{\unitlength}{1.0cm}
\begin{center}
\begin{picture}(11,10)
%\put(0.5,7.5){\LARGE a)}
%\put(0.5,5){\LARGE b)}
\put(0.0,0.0){\psfig{figure=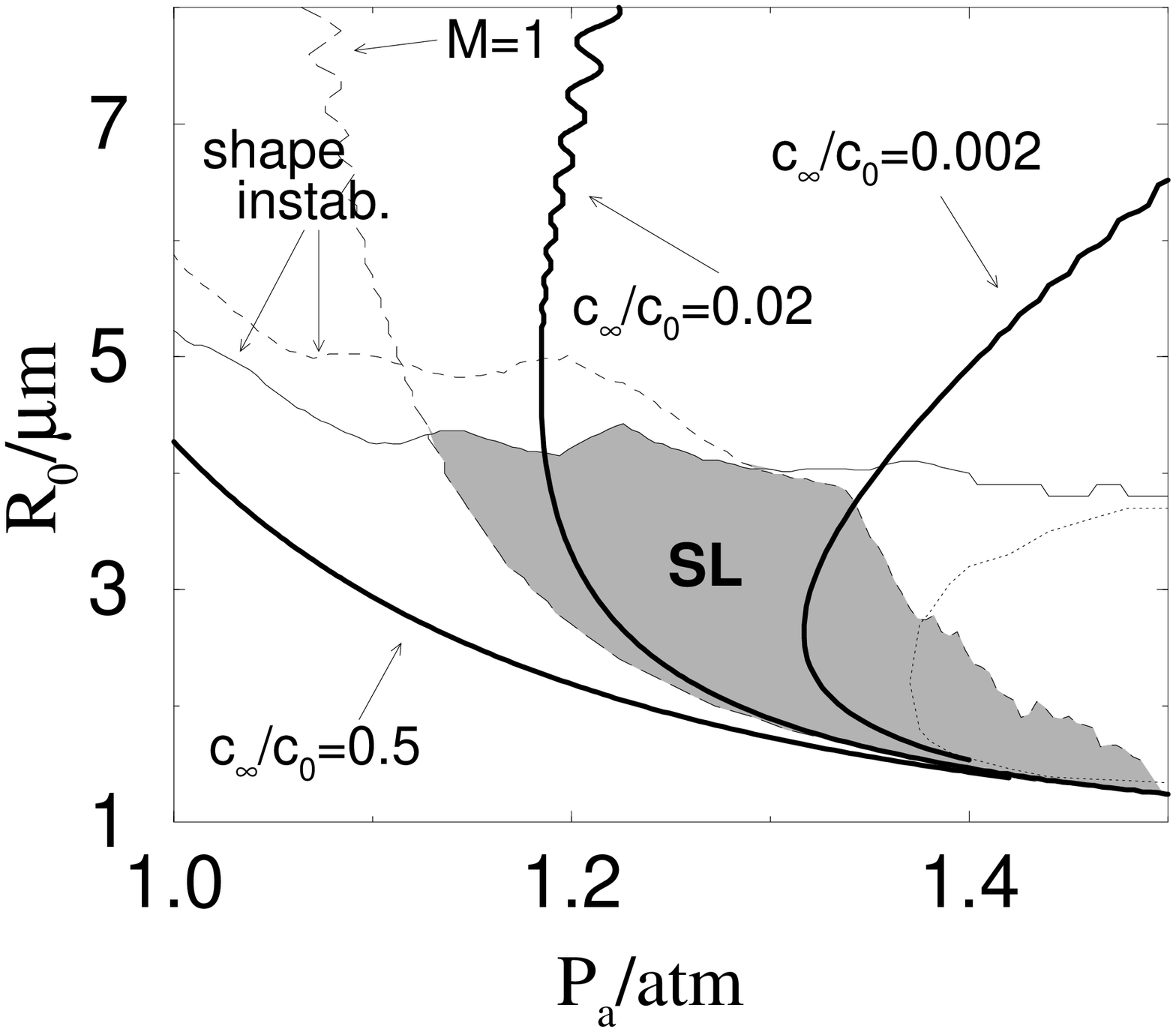,width=12cm,angle=-90.}}
%\put(0.5,0.5){\psfig{figure=dummy.eps,width=12cm}}
\end{picture}
\end{center}
\caption[]{ 
The figure shows the discussed effects for argon
all together:
Beyond the $M=1$ curve (long dashed) SL is possible.
The bubble grows thanks to rectified diffusion
right of the diffusive stability curves (shown for $c_\infty/c_0 =
0.5$, $0.02$, and $0.002$, left to right). 
At
the parametric shape instability (solid) and the
shape instability according to criterion
(\ref{rt_crit}) (dashed) micro bubbles pinch off,
but the bubble can survive in the trap for $P_a \aleq 1.4$atm.
SL is possible in the shaded region, 
stable
SL, if in addition the slope
of the diffusive equilibrium curve is positive. Here, this occurs 
for $c_\infty/c_0 = 0.002$ for $P_a \approx 1.33$atm and $R_0\approx 3\mu m$.
Right of the dotted line (criterion (\ref{adot_crit})) the bubble is thrown
out of the trap in case of a pinchoff. Consequently, if at all,
only {\it stable} SL is
possible in this high $P_a$ regime. 
}
\label{total}
\end{figure}

In order to predict which SL regime is realized for given $c_\infty$ and $P_a$,
we also have to take above shape instabilities into consideration. Therefore,
in figure \ref{total}
we plot an enlargement of figure \ref{bif_dia} together with the
thresholds for the shape instabilities 
and the energy focusing condition, taken
from figure \ref{pi_rt}. In the relevant parameter regime $P_a > 1.0 $atm
the critical $R_0^{PI}$ is about $4-5\mu m$.
Only for large $P_a \ageq 1.4 $atm
the Rayleigh-Taylor instability becomes relevant, see below for the
consequences.
As worked out in sections \ref{shapestab} and \ref{shocks},
SL can only occur in the shaded region in figure \ref{total}.
In this region {\it only for very low concentration}
$c_\infty / c_0 \sim 0.002$ {\it the bubble is diffusively stable}
and only then we can have diffusively stable SL. For larger concentration we
have diffusively unstable SL.

The different regimes of
diffusively stable and unstable SL and no SL
  can easily be read off from figure \ref{bif_dia_c}.
Take fixed $P_a = 1.1$: growing bubbles are possible for
$c_\infty / c_0 > 0.07$, if
they are large enough. They finally run into the
parametric shape instability at $\sim 5 \mu
m$ and a micro bubble will pinch off. If the remaining bubble is still large
enough, i.e., above the stability line in figure
\ref{bif_dia_c}, the process 
will repeat. The allowed size of the bubble after the pinch off is very much
restricted; for $c_\infty/ c_0 = 0.1$ it is only in the $R_0$ range between
$4.6\mu m$ and $5\mu m$. If the pinched off micro bubble is larger, the
remaining bubble will dissolve. No diffusively stable regime exists.

For $P_a = 1.3$ the situation is quite different. For concentrations in a window
$0.003 < c_\infty/c_0 < 0.005$ bubbles in the $R_0$ regime $\sim 2 \mu m$ up to
$5\mu m$ will grow and {\it stabilize}. As in this $P_a$ regime the
energy focusing
criterion is fulfilled, we have stable SL. Here we are at the very core of
Barber's finding \cite{bar95} that stable SL is only possible for low
$c_\infty$ and high $P_a$.  
If the relative concentration is even smaller than $0.003$,
bubbles will dissolve and no SL is possible. For larger concentration large
enough bubbles $\ageq 1.8 \mu m$ (almost independent of $c_\infty$) will
grow up to $\sim
5\mu m$ where micro bubbles pinch off. In contrast to smaller
$P_a\sim 1.1$atm,
the remaining bubble
is very likely to end up in the (now much larger) growing $R_0$ regime.
Pinched off micro bubbles with $R_0 < 1.8 \mu m$ will dissolve.

The same domains no SL, unstable SL, and stable
SL can be identified from
figure \ref{bif_dia} or figure \ref{total};
also these plots make it very evident  that high
$P_a/P_0$ and low $c_\infty/c_0$ are necessary to obtain stable SL.

The total phase diagram in the $c_\infty $ -- $P_a$ phase space, our main
result, was already presented in
figure \ref{phase_dia}. The notation in that diagram is as follows: If we denote
a regime with stable SL or unstable SL we mean that there are bubbles of
certain radius
which are diffusively stable or growing, respectively; other, smaller bubbles
dissolve. For $P_a \aleq 1.17$ the $c_\infty/c_0$ window of stable SL shrinks to
zero. For very low $c_\infty$ the no-bubble regime is very extended. If one
now slightly increases $c_\infty$, one immediately enters a regime where
the
energy focusing
condition is fulfilled. This may be the prime reason why it is so
much easier to find SL, diffusively stable or not,  for low concentration.

The experimental observations of fig.\ 4 in ref.\ \cite{bar95} are in agreement
with our analysis.
For that figure $P_a$ is in the range of 1.3atm. Then we have
unstable SL for large concentrations $c_\infty/c_0 = 26\%$ and $6.6\%$ and
stable SL for low
concentration $c_\infty/c_0 = 0.4\%$.

What happens for very large forcing amplitudes $P_a$?
From figure \ref{total} we see that the SL regime (shaded) intersects with
the regime where bubbles cannot survive the micro bubble pinchoff (right
of the dotted line (\ref{adot_crit})
in figure \ref{total}) because of the violent
short time scale Rayleigh-Taylor instability.
Consequently, in this high $P_a$
regime only {\it stable} SL should be possible,
a prediction which is worth while being tested experimentally.
Extremely low argon concentrations
$c_\infty / c_0 \aleq 0.0006$ are necessary to guarantee
diffusively stable equilibria in this regime. 
According to the approximate curve (\ref{adot_crit}) of the onset of the
RT instability in figure \ref{total}, this regime starts
at about $P_a\approx 1.4$atm, but as pointed out in
subsection II-D, this number should not be taken too strictly. 
For lucidity we
did not put in the right borderline of the unstable SL regime in
the phase diagram figure \ref{phase_dia}. -- The upper borderline of
the unstable SL regime is discussed in subsection IV-E.

In our former publication \cite{bre96}
we have stressed the importance of wiggles such as
those in the graphs in figures \ref{pi_rt} -- \ref{total}.
As a consequence,
for large enough $P_a$ and small enough $R_0$ one may have
stable SL with different, discrete $R_0$; i.e., {\it multiple} stable
equilibria.
We will present some results of full numerical simulations on this issue in
section \ref{compare}.

Possibly an experimental hint to multiple stable equilibria has been found by
Crum and Cordry \cite{cru94b}.
After registering SL light from a bubble for a few seconds, 
they distorted the bubble and observed the SL intensity
to jump from one constant value to another (smaller) one. They
suggested that the discrete light intensities corresponded to discrete
diffusively stable radii.

For the dynamics chosen here the multiple stable equilibria
are just parametrically unstable as they lie beyond $R_0\approx 5\mu m$.
This should not be taken too strictly, as a slight change of the
model
may allow for observable multiple stable equilibria.
However, we  see that
the wiggles are {not necessary} for stabilization. An increase of
$\la p \ra_{t,4}(R_0)$ with $R_0$ is sufficient, i.e., $\beta > 0$. 
If there are wiggles, we in addition have discretization of the
equilibria.

\subsection{Scaling laws}
The average slope and the wiggles in figure \ref{p_tau}
can be understood in detail as shown in ref.\ \cite{bre96c}.
Here we only quote scaling laws
for the running
average of $\la p \ra_{t,4} (R_0) $ which smooths out any wiggles, 
\begineq
\overline{\la p\ra}_{t,4} = 
 {1\over 2\Delta R_0 }
\int_{R_0 -\Delta R_0}^{R_0+\Delta R_0}
 \la p\ra_{t,4}
 (R_0 + x ) dx 
\label{running_av}
\endeq
with $\Delta R_0 \geq 0.5\mu m$.
For very small $R_0$ the surface tension term dominates in the RP equation
(\ref{rp}) and 
$ \la p\ra_{t,4}\propto R_0^{-1}$.
For slightly larger $R_0 <R_0^{crit}$ we have  
$ \la p\ra_{t,4}\propto R_0^{-3/2}$. For large $R_0 > R_0^{crit}$ (where
wiggles
occur and the running average becomes necessary) the first and the second
term on the rhs of (\ref{rp}) balance and
\begineq
\overline{\la p\ra}_{t,4}\propto R_0^{6/5},
\label{sca1}
\endeq
i.e., the average slope
$\beta$ is positive and equilibria are stable. The critical ambient
radius $R_0^{crit}$ beyond which wiggles occur and stability is achieved
scales like \cite{bre96c}
\begineq
R_0^{crit} \propto {1\over P_a - P_0 },
\label{sca2}
\endeq
the corresponding average pressure like
\begineq
\overline{\la p^{crit}\ra}_{t,4}\propto
\left( {1\over P_a -P_0 }\right)^{9/2}.
\label{sca3}
\endeq

The scaling law (\ref{sca1})
directly reflects the average slope $R_0 \propto
c_\infty^{5/6} $ in figure~\ref{bif_dia_c}. 
Thus for large enough $R_0$ there is always diffusive stabilization,
but only for low $c_\infty$ and high $P_a$ this will be in a parametrically
stable domain where the energy focusing condition is fulfilled.

A detailed discussion of these types of scaling laws will be presented
elsewhere \cite{bre96c}.

\subsection{Diffusively unstable SL bubbles}
In the previous subsections
 we saw that our theoretical stability diagram is in
agreement with experiment. What about the growth rates in the unstable SL
regime where
the bubble grows by rectified diffusion and finally hits the parametric
instability line at about $R_0 = 5\mu m$?
Because of the growth (i) the {\it relative phase} of light emission $\phi
(R_0) $ will slightly change, (ii) the light intensity will increase as more
and more gas is in the bubble, and (iii) the maximal radius will increase.
When the shape instability line is hit, a micro bubble pinches off,
giving the bubble a recoil. As this repeats again and again
on the diffusive time scale of $\sim 1s$, the bubble
seems to ``dance''.

Within the adiabatic
approximation we now calculate $\phi (R_0 (\bar t )) = \phi
(\bar t)$. According to
the energy focusing mechanisms discussed in section \ref{shocks}
light can be  emitted if the
(inward) bubble wall velocity becomes supersonic, 
$M = {- \dot R /  c_{gas}} \ageq 1$.
We define the time when $M=1$ holds as $t_s$.
The relative shift of this time to the forcing phase defines $\phi_s$.
As the
waves and shock waves in the bubble are
very fast, we take the time $ t_s$ of the
detachment of the (shock) wave as the time of the light pulse.
The error we make by this
approximation is of the order of 
$\Delta t \sim
R_0/c_{gas} \sim 1 ns$, as follows from a simple estimation. 
%As the velocity of sound $c_{gas}$ we take that
%of an ideal monatomic gas. For an isothermal gas this 
%is the ambient speed
%of sound in the gas $c_0 \sim 300 m/s$, for an adiabatic 
%$c_{gas} = c_0 R_0/R $.
%The shock  detaches when $R(t)$ is smaller than $R_0$ and larger than
%the van der Waals hard core value $R_0/8.86$.

Next we calculate $R_0 (\bar t )$ from eq.\ (\ref{fsgrowth}) for discrete
adiabatic times $\bar t = n T$. Take fixed $R_0$ and calculate
the time averages
$\la . \ra_{t,i}$ from the RP equation. The integral
\begineq
I= \int_0^\infty {dh' \over \la ( 3h' + R^3(t))^{4/3}\ra_{t,0}}
\label{int}
\endeq
can be calculated numerically. Its convergence, however, is slow.
In order to speed up the numerical calculation, 
for calculations over a long period of time 
we approximate the integral $I$ by
$I= a/ R_{max} + (1-a)/ R_0 +
(3h_{max})^{-1/3}$.
Here,
$a\approx 0.9$ is an adjustable parameter, which slightly depends
on $R_0$ and $P_a$, and $h_{max}\gg R_{max}^3$
must be sufficiently large. The approximation
is very well controlled and the results are indistinguishable from
the exact result. 
The growth during the time interval $T$ finally 
reads
\begineq
\Delta R_0 (\bar t ) = {TDc_0\over \rho_0 R_0^2(\bar t ) I }
\left[ {c_\infty \over c_0} - {\la p (t)\ra_{t,4} (\bar t )\over P_0}\right].
\label{delr0}
\endeq
$\Delta R_0$ is added to $R_0$ and the procedure is repeated until $R_0(\bar t
)$ hits the parametric instability curve.
Here from figure\ \ref{pi_rt} we took $R_0^{PI} = 5\mu m$ as a very
good approximation. 
A random fraction of the
bubble will pinch off.
In figure \ref{phi_fig}  we show
the ambient radius $R_0 (\bar t ) $ and the relative phase
of the light pulse $\phi_s (\bar t ) $ for three different
relative gas concentrations $c_\infty/c_0$.

Figure \ref{phi_fig} should be compared to the corresponding
experimental figure 4 of ref.\ \cite{bar95}. Unfortunately, for that figure
the precise forcing pressure amplitude $P_a$ and the ambient bubble
size are not known.
So we try both $P_a=1.3$atm (fig.\ \ref{phi_fig}a) and
$P_a=1.2$atm (fig.\ \ref{phi_fig}b).
For $P_a=1.3$atm we have very good agreement with experiment:
stable SL for 3mmHg, growth for 50mmHg and for
200mmHg. For $P_a = 1.2$atm a bubble in a fluid with 3mmHg argon
concentration would dissolve according to our approximation,
see figure~\ref{phase_dia}, so we choose 12mmHg as smallest concentration
and find stable SL.

Let us compare the {\it growth rates} with the experimental values.
As in experiment, for larger gas concentration
the growth rates of $R_0$ and $\phi_s$ strongly increase.
Quantitative estimates from figure \ref{phi_fig}
for the growth rates of the phases for
both forcing pressures and both concentrations are compared
to the experimental ones in table 1. They are slightly larger from what
is found in experiment, but agree in order of magnitude. 
%Note that such a comparison in principle allows for an indirect
%measurement of the radius. 
Rather than focusing on an exact quantitative agreement
here, the important point is that for lower concentration
$c_\infty/c_0 = 0.0165$ (corresponding to 3mmHg)
the phase of light emission is {\it stable}
due to the diffusively stable ambient
radius (for $P_a=1.3$atm).
In this example, the bubble is locked at
$R_0=4.31\mu m$, and the phase of the light pulse is stable for ever.

\begin{figure}[p]
\setlength{\unitlength}{1.0cm}
\begin{center}
\begin{picture}(11,20)
\put(-1.0,17.5){\LARGE a)}
\put(-1.0,8){\LARGE b)}
\put(0.0,0.5){\psfig{figure=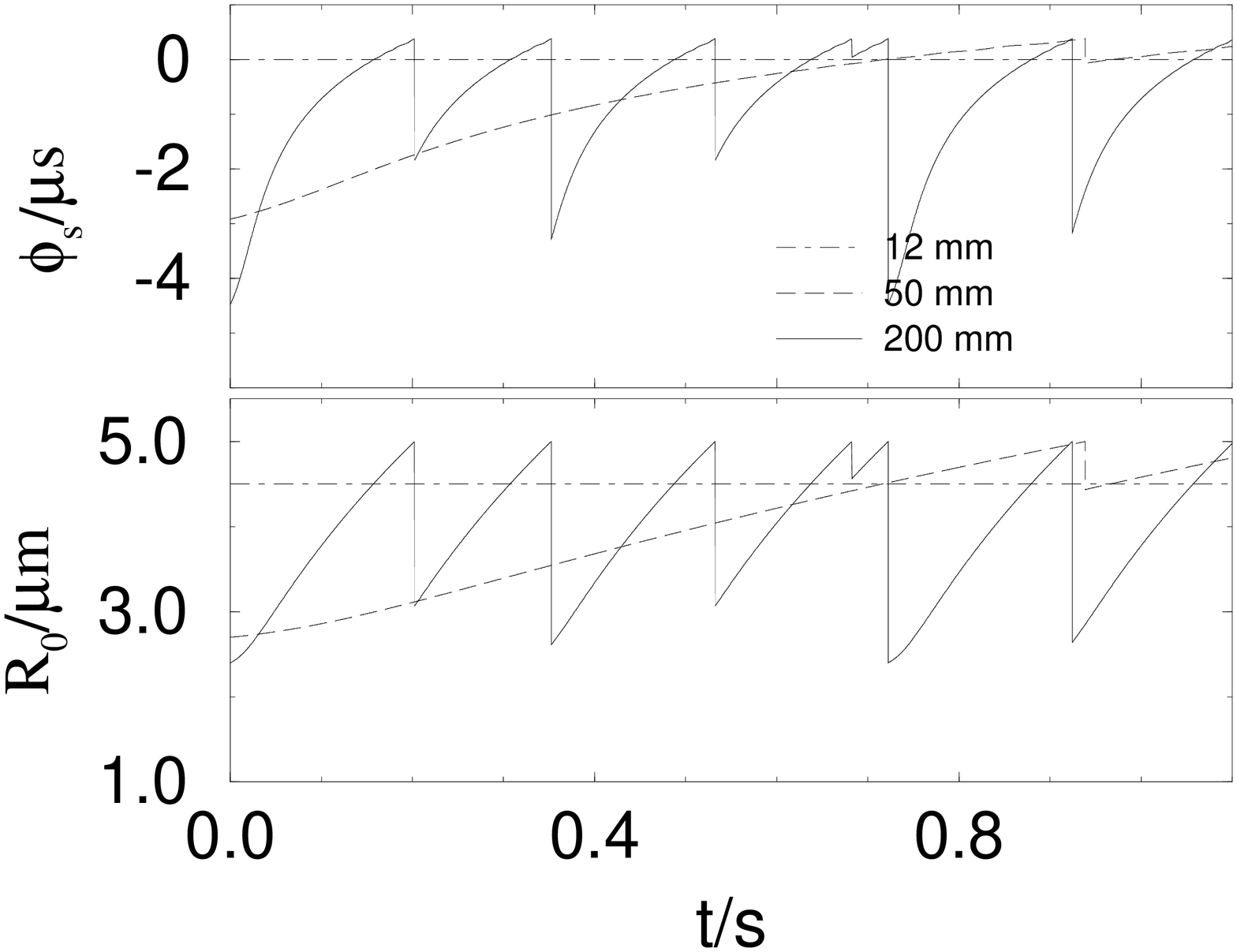,width=11cm,angle=-90.}}
\put(0.0,10){\psfig{figure=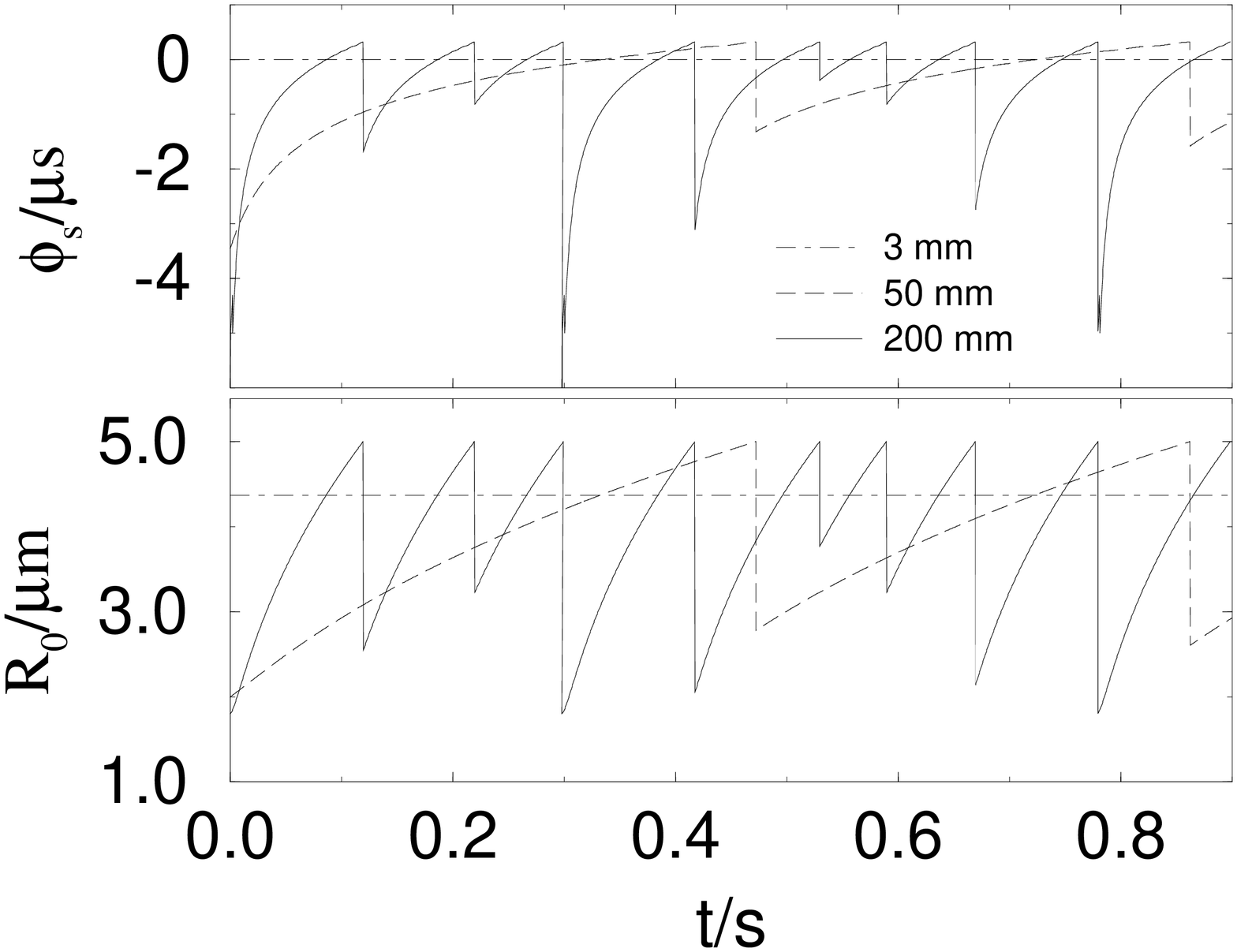,width=11cm,angle=-90.}}
%\put(0.0,0.5){\psfig{figure=dummy.eps,width=11cm}}
%\put(0.0,10){\psfig{figure=dummy.eps,width=11cm}}
\end{picture}
\newpage
\begin{picture}(11,11)
\put(-1.0,9){\LARGE c)}
%\put(0.0,0.0){\psfig{figure=figure11_c.ps,width=11cm,angle=0.6}}
\put(0.0,0.0){\psfig{figure=dummy.eps,width=11cm}}
\end{picture}
\end{center}
\caption[]{
The phase of the light pulse $\phi_{s}(\bar t)$ (upper)
and the corresponding
ambient radius $R_0(\bar t )$  (lower) for (a) $P_a= 1.3$atm and
for (b) $P_a=1.2$atm,  for three different gas concentrations
$c_\infty / c_0 =  0.00395$,
$c_\infty / c_0 =  0.0658$, and 
$c_\infty / c_0 =  0.26$, corresponding to a gas pressure of
$3$mmHg, $50$mmHg, and $200$mmHg, respectively. These values are
chosen as in experiment (figure \ref{phi_fig}c)
to which these figures compare very well. 
For $P_a=1.2$atm we had to choose 12mmHg as smallest concentration,
as for 3mmHg the bubble would still dissolve.
Diffusively stable SL is only seen for the lowest concentration.
The strength of the micro-bubble pinch-off at $R_0=5\mu m$, i.e.
the decrease of the ambient radius, is chosen randomly.
(c) Experimental result for the phases of light emission
for the same three gas concentrations as in (a). This figure is reproduced
from figure 4 of Barber et al.\ \cite{bar95} with kind
permission by the authors. It also shows the
relative phase of light emission for air bubbles: Stable SL is achieved
for much higher concentration $c_{\infty}^{air}/c_0 = 0.2$, corresponding
to 150mmHg. The discrepancy between air and argon can be resolved by
also considering chemical instabilities \cite{loh96,bre96b}.
}
\label{phi_fig}
\end{figure}

What is the physical consequence of the large bubble growth rates
obtained for large argon concentration $c_\infty/c_0$ (table 1)?
The shape instability line will be hit more frequently per unit time
and the bubble's dancing frequency
will thus become larger, as more micro bubble
pinch-offs and resulting bubble recoils
will take place per unit time.
As noted above with some certain probability
{\it per pinch-off} the pinched off micro bubble(s)
are too large so that the remaining bubble dissolves.
Thus with increased pinch off frequency this probability {\it per time}
increases.  We speculate that this mechanism sets the upper threshold
of the unstable SL regime (towards a no SL regime) in the phase diagram 
figure \ref{phase_dia}.
Table 1 teaches us that the growth rates increase
drastically with 
$c_\infty/c_0$, so the probability of having a long living bubble
for high gas concentration becomes very low.
Indeed, Gompf \cite{gom96} reported that the larger the concentration is
for fixed $P_a$, the faster the unstable SL bubble dies. If $c_\infty$ is big
enough, it will thus be very unlikely for the bubble to survive an
appreciable time.
From experiment \cite{bar95} we
know that the upper concentration
threshold of unstable SL is beyond $c_\infty/c_0 = 0.26$.

This hypothesis also explains why the water in the SL container
``ages'' \cite{gom96}, 
in case the container is not gas tight.
By ``aging'' it is  meant that stable SL and finally also unstable
SL becomes impossible with ``old'' water.
The reason is that 
external air
diffuses into the water,
dissolves, and $c_\infty/c_0$ increases. Consequently,
the originally stable bubble
is pushed into the unstable regime and starts to ``dance'', shedding
off micro bubbles. The dancing frequency becomes higher and higher 
and finally the bubble dissolves
after a too large pinch off.
Bubbles may be reseeded, but will also die very soon.

What we do not understand in the unstable regime
is the dependence of the light intensity
on the gas concentration as e.g.\ measured in figure 2 of \cite{wen95}.
We speculate that it depends on the ambient size of the bubble which
is supported by figure 6 of ref.\ \cite{loe95}. In that figure
L\"ofstedt et al.\ show that the maximal radius and the SL intensity
are correlated.

\vspace{1cm}

% \begin{table}[htp]
 \begin{center}
 \begin{tabular}{|c|c|c|c|c|}
 \hline
         $  $ 
       &  $c_\infty  $ 
       & $P_a=1.2$atm
       & $P_a = 1.3$atm
       & experiment \cite{bar95}
 \\
\hline
         $ \phi_s $ 
       & $50$mmHg
       & $1.7 \mu s/s$
       & $2.7\mu s/s$
       & $0.5\mu s/s$
\\
         $  $ 
       & $200$mmHg
       & $9\mu s/s$
       & $11\mu s/s$
       & $5\mu s/s$
\\
\hline
         $ R_0 $ 
       & $50$mmHg
       & $2.4 \mu m/s$
       & $4.7\mu m/s$
       & ---
\\
         $  $ 
       & $200$mmHg
       & $11 \mu m/s$
       & $18\mu m/s$
       & ---
\\
 \hline
 \end{tabular}
 \end{center}
% \end{table}

\vspace{0.5cm}

{\it 
\centerline{\bf Table 1} \noindent
Growth rates for the phase of light emission $\phi_s$ for
two forcing pressure amplitudes $P_a$ and two concentrations
$c_\infty$
near the pinch off of the micro bubble in comparison
with the experimental data \cite{bar95} for which the forcing is not
exactly known.
Stronger forcing and larger argon
concentration enhance the growth. We find order
of magnitude agreement. We also give the growth rates
for the radii which are not experimentally available.
}

\vspace{1cm}

Finally we address the question whether
the growth of $\phi_s$ has to be monotonous. In fact, it does
not.
For larger $R_0$  ($\approx 6\mu m$, where the bubble is already
parametrically unstable)
the growth rate of $\phi_s$ is  {\it wiggly}.
Similar oscillations show up in the maximal radius
$R_{\rm max}(t) = {\rm max} \{R(t')| t\le t'\le t+T \}$,
as e.g.\ seen in figure 4 of Barber et al.\ \cite{bar94}.
This wiggly structure as a function of
{\it time} is
a direct consequence of the wiggly structure
of $\phi_s$ and $R_{max}$ as a function of {\it ambient radius} $R_0$
which is due to a resonance phenomenon in the RP equation \cite{bre96c}.
The growing $R_0$ probes the wiggles in
$\phi_s (R_0)$ and $R_{max}(R_0)$.

In experiment no or hardly any oscillatory structure in $\phi_s$ is seen
\cite{bar95}. Consequently, the ambient radius
  $R_0$ does not seem to be in the wiggly regime.
Thus multiple stable diffusive equilibria may only be important
in the shape unstable regime in the $R_0-P_a$ parameter space.
Indeed, the wiggly structure in $R_{max}(R_0(\bar t))$ \cite{bar94}
is only revealed when
boosting the bubble in the unstable SL regime where it
becomes shape unstable after a few ms and bursts
(cf. fig.\ 4 of \cite{bar94}).

\begin{figure}[htb]
\setlength{\unitlength}{1.0cm}
\begin{center}
\begin{picture}(11,10)
%\put(0.5,7.5){\LARGE a)}
%\put(0.5,5){\LARGE b)}
\put(0.0,0.0){\psfig{figure=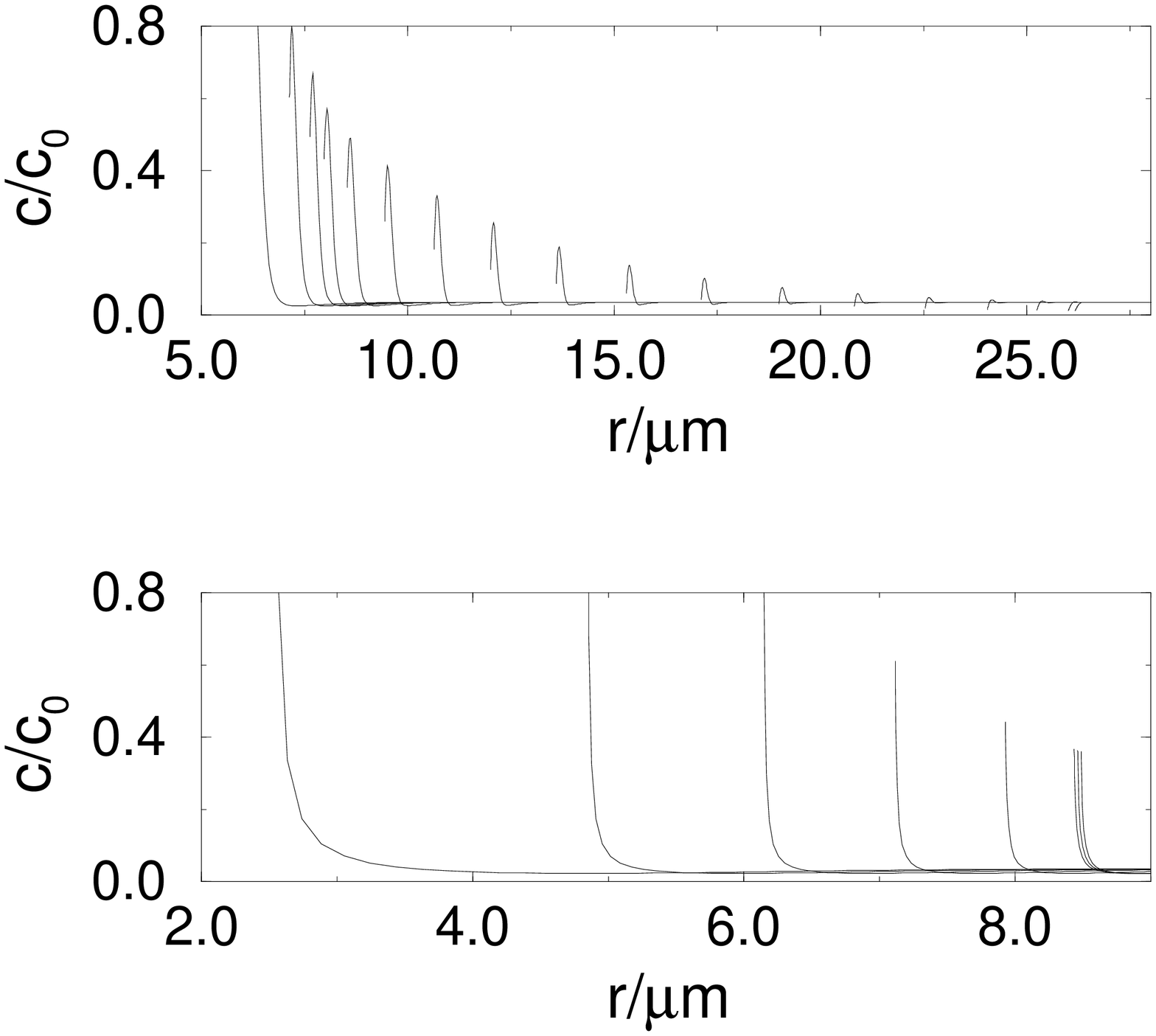,width=12cm,angle=-90.}}
%\put(0.,0.){\psfig{figure=dummy.eps,width=12cm}}
\end{picture}
\end{center}
\caption[]{
Gas concentration profiles outside the bubble for expanding (a) and
collapsing (b) bubble radius, respectively. The bubble is near diffusive
equilibrium, driving pressure is $P_a=1.15$atm. Profiles are shown at
intervals of $0.75\mu s$ (a) and $10ns$ (b).}
\label{profiles}
\end{figure}

\section{Comparison of the adiabatic approximation 
to the full numerical solution}\label{compare}
\setcounter{equation}{0}
\subsection{Concentration profiles}
To compare our results within the adiabatic approximation with 
the exact solution, we
must numerically solve the PDE (\ref{convdiff}) with the 
boundary
conditions given above. We sketch our numerical method 
in appendix \ref{appa}.
Here we report on results.
We do not consider shape instabilities in this section.

\begin{figure}[htb]
\setlength{\unitlength}{1.0cm}
\begin{center}
\begin{picture}(11,10.6)
%\put(0.5,7.5){\LARGE a)}
%\put(0.5,5){\LARGE b)}
\put(0.0,0.0){\psfig{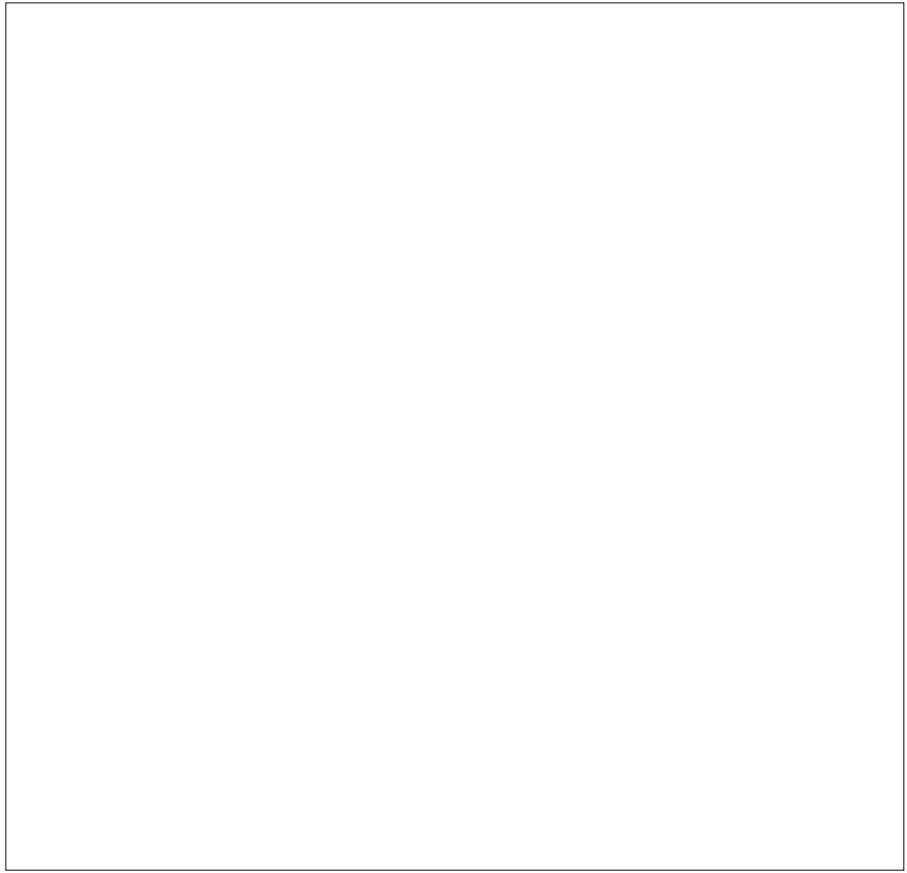}}
%\put(0.0,0.0){\psfig{figure=figure13.eps,width=12cm}}
\end{picture}
\end{center}
\caption[]{  
Three different stable equilibria are approached both from
above and from below. Shrinking or growth are hardly noticable.
The six initial radii are $6.035\mu m,6.05\mu m$; $6.32\mu m,6.34\mu m$;
$6.65\mu m$ and $6.665\mu m$,
respectively. Otherwise, all conditions
are the same. Again, we choose $P_a=1.15$atm.
}
\label{r0_long}
\end{figure}

In fig.\ \ref{profiles} we show concentration profiles of the gas
outside the bubble during expansion and collapse. We take an argon
bubble driven at
$1.15$atm
close to diffusive equilibrium. Near the minimum the
concentration gradient at the bubble wall is negative 
and the bubble ejects
gas which accumulates near the bubble wall as the 
diffusive time scale is slow
compared to the bubble motion. When the bubble is 
reexpanding, it pushes away
the accumulated gas together with the fluid. During the 
expansion phase the
concentration $c(R(t),t)$ at the bubble wall decreases 
due to Henry's law
(\ref{henry}). At some point the gradient becomes 
positive and the mass content
of the bubble grows.
The wall of gas outside the bubble is thus (i) pushed 
away from the bubble,
(ii) deaccumulates (towards smaller $r$) because of 
bubble growth, and (iii)
shrinks because of diffusion (towards larger $r$).

Apart from the diffusive processes in the fluid these profiles 
only mirror the
time and space dependence of $c_{osc}(r,t)$.
The width of the boundary layer in which $c_{osc}$
is a dominant feature of the concentration profile can
be readily estimated as
\begineq
\delta_D = \sqrt{D\over\omega} ,
\label{deltad}
\endeq
which is the characteristic length for diffusive processes
on a time scale $\sim T$.

\begin{figure}[htb]
\setlength{\unitlength}{1.0cm}
\begin{center}
\begin{picture}(11,10)
%\put(0.5,7.5){\LARGE a)}
%\put(0.5,5){\LARGE b)}
\put(0.0,0.0){\psfig{figure=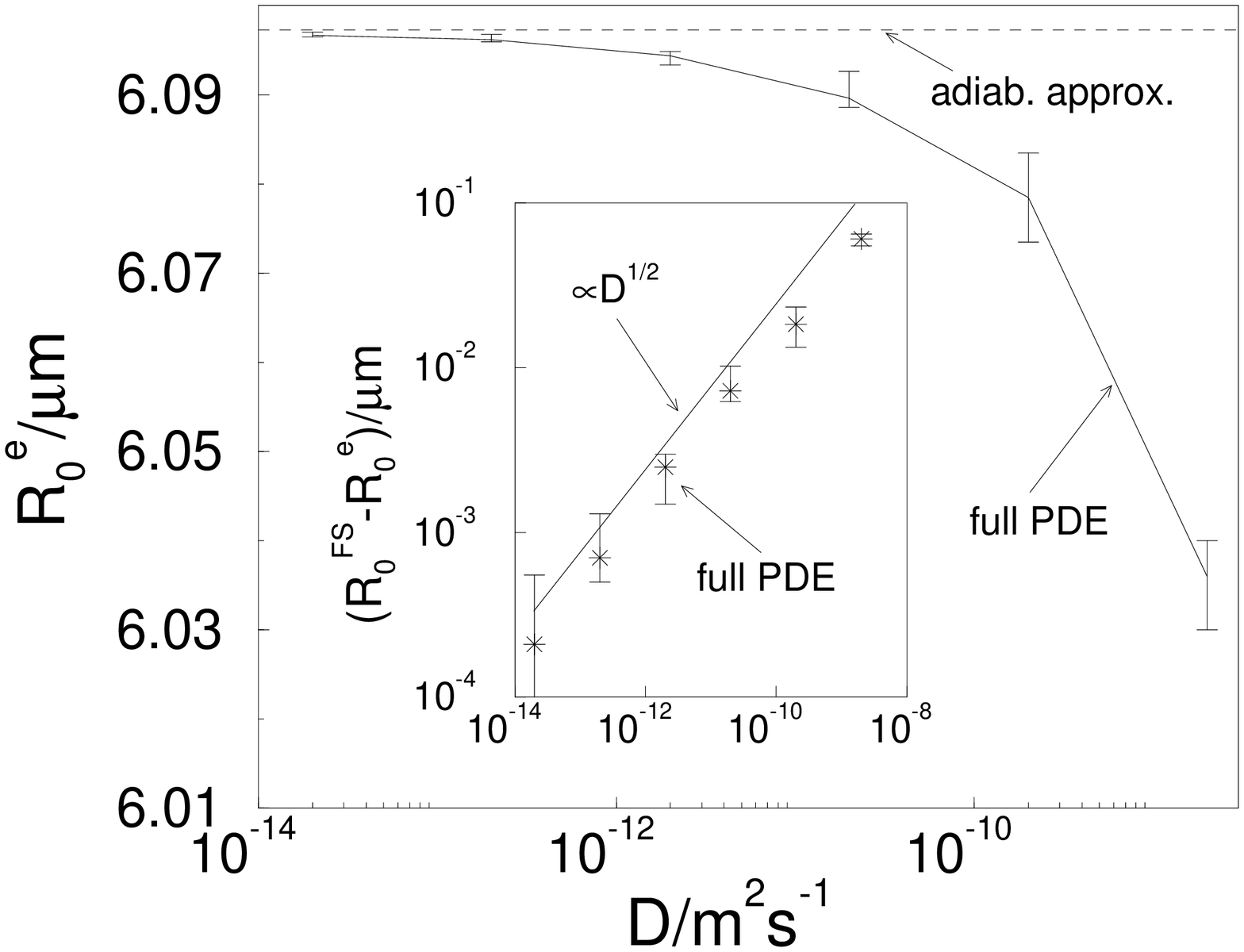,width=12cm,angle=-90.}}
%\put(0.,0.){\psfig{figure=dummy.eps,width=12cm}}
\end{picture}
\end{center}
\caption[]{ 
Location of an equilibrium radius for various $D$ (solid line).
At the upper end of the error bar the bubble is shrinking,
at the lower end it is growing.
The theoretical
value from adiabatic approximation is indicated by the dashed line.
Note that the tiny difference
of $\approx 60$nm (for the physical diffusion
coefficient $D=2\cdot 10^{-9}m^2/s$)
is not experimentally detectable at the moment.
The bubble is driven at
$1.15$atm with a gas saturation in water of $c_{\infty}/c_0 = 0.035$.
The inset shows the deviation $R_0^{adiab}-R_0^e$ drawn in a log-log diagram,
together with our estimate (\ref{dev}) (solid line).}
\label{approach}
\end{figure}

\subsection{Comparison on diffusive equilibria}
First, let us focus on the equilibrium radii for diffusively stable
bubbles.  
Can we find the multiple equilibria predicted in section \ref{diffstab} 
in the full numerical
simulation, i.e. different stable situations for the same physical
parameters?
Figure \ref{r0_long} shows $R_0(t)$ for 
several different initial ambient 
radii. The diffusion constant has the physical value 
$D=2\cdot 10^{-9}m^2/s$, corresponding to a Schmidt
number of $Sc\approx 3000$ in the
regime of interest ($\omega\approx 2\pi\cdot 26.5 kHz, R_0\approx 6\mu m$).
We indeed observe several stable and unstable 
equilibria. However, they
deviate slightly from those calculated in the adiabatic
approximation.

The deviation, however, is tiny, less than $0.06\mu m$ as seen from figure
\ref{approach} and clearly not detectable with today's experimental
possibilities. Thus for all practical reasons we can consider the adiabatic
approximation of the equilibrium radii as exact.

Nevertheless, let us wonder what the origin of the deviation is. It 
can be explained by considering higher 
order corrections to the adiabatic theory \cite{fyr95}: the equilibrium
condition (\ref{equi}) is
modified, lowering the required mean pressure at the 
bubble surface for stability to
\begineq
{\left< p(t) \right>_{t,4}\over P_0} = {c_\infty\over 
c_0} - {1\over Sc^{1/2}}\cdot{4\over c_0}\left< {1\over R^3}
\int_0^\infty c_{osc}^0(\tilde{h},t)d\tilde{h} \right>_{t,4} ,
\label{equi2}
\endeq
$c_{osc}^0$ being the zeroth order (in $Sc^{-1/2}$) solution of
the oscillatory part of the profile, depending on
$\tilde{h}\equiv Sc^{1/2} h$ \cite{fyr95}. The shift in
$\left< p(t) \right>_{t,4}$ causes a corresponding shift
$R_0^{adiab}-R_0^e$ in the equilibrium radius $R_0^e$.

To analyze the deviations from the equilibrium position
(\ref{equi}) further, 
we redid the numerical PDE calculation for smaller
(unphysical) $D$. As expected, in the $D\to 0$ limit the 
($D$ independent) adiabatic 
fixed point is approached as can be seen from fig. 
\ref{approach}. For that
figure the numerical equilibrium radii are ``measured''
by detecting either ``shrinking'' or ``growth'' for
slightly different radii.
The stable equilibrium is determined by linear interpolation between the
growth rates of a growing and a shrinking bubble. In principle, 
we can achieve
arbitrarily small error bars 
in our numerical results. This is confirmed by the 
excellent agreement of the $D\to 0$ limit with the adiabatic approximation.

There is a theoretical possibility that the deviation of
the full PDE dynamics from the adiabatic approximation matters,
namely when multiple stable equilibria
are to be resolved. For completeness
we  discuss this point in Appendix \ref{appb}.

\begin{figure}[htb]
\setlength{\unitlength}{1.0cm}
\begin{center}
\begin{picture}(11,10)
%\put(0.5,7.5){\LARGE a)}
%\put(0.5,5){\LARGE b)}
\put(0.0,0.0){\psfig{figure=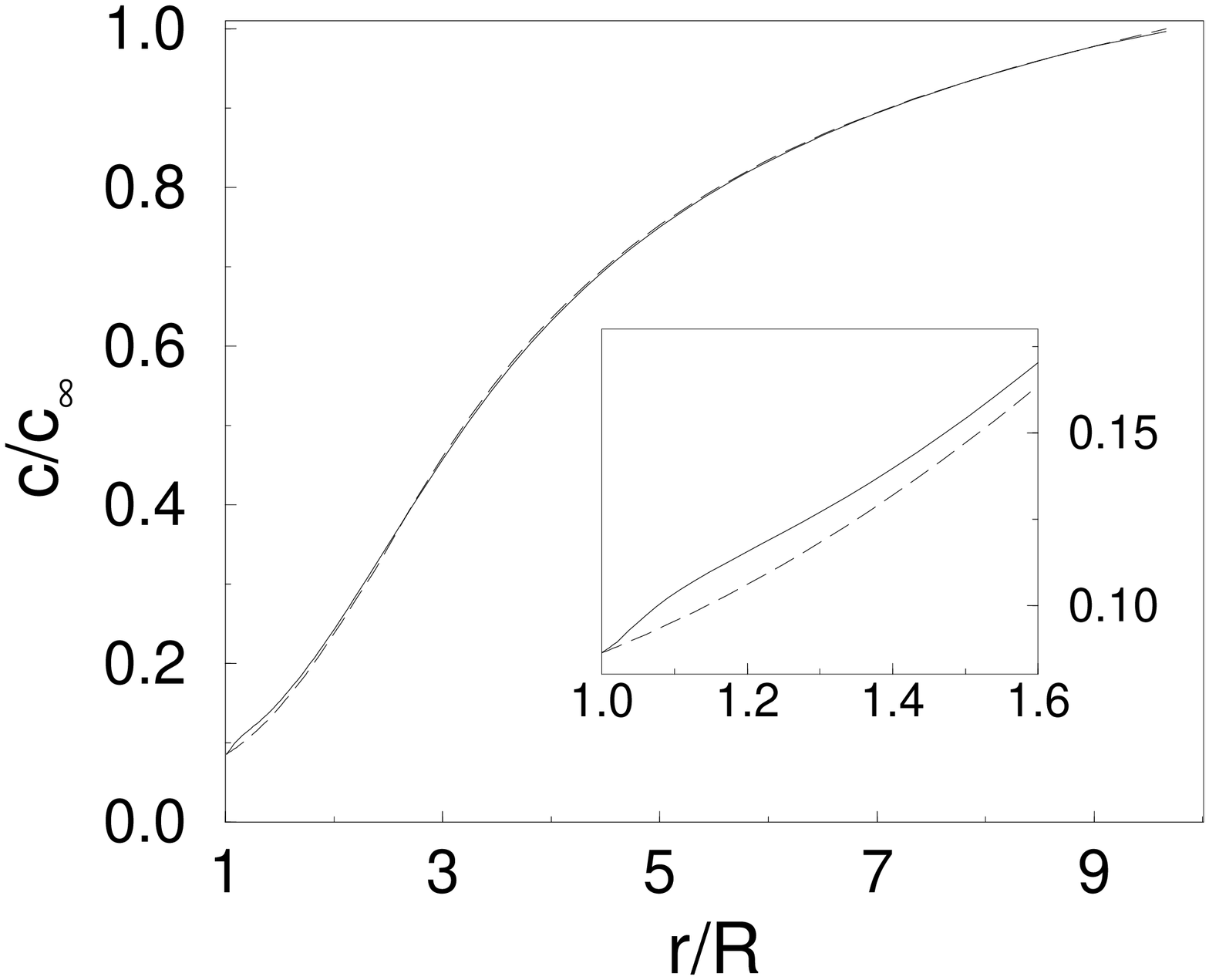,width=12cm,angle=-90.}}
%\put(0.5,0.5){\psfig{figure=dummy.eps,width=12cm}}
\end{picture}
\end{center}
\caption[]{ 
Comparison of $\langle c(r,t)\rangle_{t,4}$ (solid line) and the
smooth profile $c_{smo}(r,t)$ from the adiabatic
approximation (dashed). The bubble ambient
radius shows pronounced growth in this case. The inset shows an enlargement
of the small $r$ regime. }
\label{cfprof}
\end{figure}

\subsection{Comparison on growth rates and profiles}

To detect the location of an equilibrium point along the 
$R_0$ axis,
starting with a constant gas concentration profile $c(h) 
= c_\infty$ is
obviously a good choice, because $\bar{c}_{smo}$ will approach 
that value
for all $h$ in equilibrium (apart from higher order 
corrections). The
sign of $\dot{R}_0(t)$ will be correct after a small 
number of oscillation
cycles.
If one is, however, interested in the actual value of 
the bubble growth rate,
i.e. bubble dynamics far from equilibrium points, 
choosing $\bar{c}_{smo}(h)$ as
initial concentration profile will avoid transients on 
diffusive time scales.
Indeed, 
\begineq
\la c(h,t)\ra_{t,4} \to \bar c_{smo} (h)
\label{cav}
\endeq
holds to very good accuracy for such a calculation 
(see figure \ref{cfprof}).
Moreover, the observed growth rate $\dot{R}_0(t)$ is in 
very good agreement
with the value calculated from (\ref{fsgrowth}), as 
indicated in figure \ref{r0_short}.

\begin{figure}[htb]
\setlength{\unitlength}{1.0cm}
\begin{center}
\begin{picture}(11,10)
%\put(0.5,7.5){\LARGE a)}
%\put(0.5,5){\LARGE b)}
%\put(0.0,0.0){\psfig{figure=figure16.eps,width=12cm,angle=-90.}}
\put(0.0,0.0){\psfig{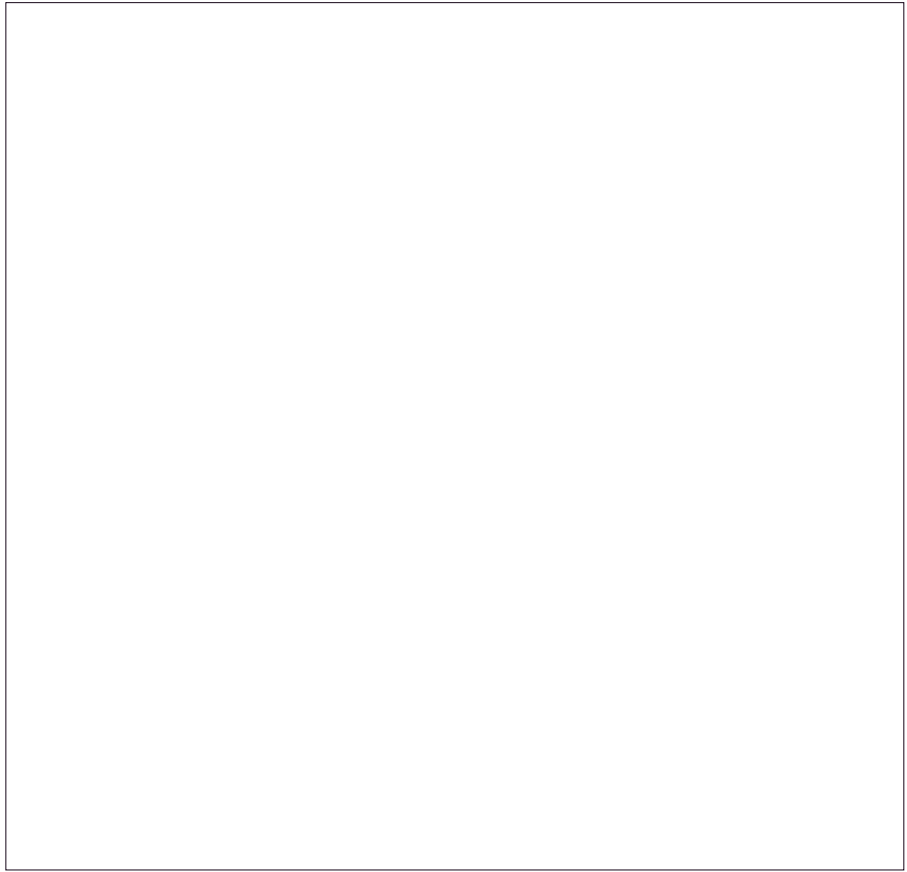}}
\end{picture}
\end{center}
\caption[]{  
$R_0(t)$ for $c_\infty/c_0 = 0.001$ and for 
 $c_\infty/c_0 = 0.3$ resulting in shrinking and growth of
 the bubble respectively. 
We chose $P_a=1.15$atm and a bubble with $R_0(t=0)=5.5\mu m$.
The adiabatic approximations are also shown as straight lines.
By definition they do not follow the violent mass exchange processes
during one cycle of time $T$. 
}
\label{r0_short}
\end{figure}

\section{Conclusions}\label{concl}
\setcounter{equation}{0}
\noindent
We worked out a hydrodynamic approach towards SL, based on the
Rayleigh-Plesset equation. This allows us to explore a considerable
part of the phase space and to study long term dynamics. As necessary
conditions for SL we demanded shape stability and energy focusing.
The adiabatic approximation \cite{fyr94,loe95}
allows us to study also diffusive
stability within the RP approach.

We have presented phase diagrams in the $c_\infty - P_a$, $R_0 - P_a$,
and $R_0 - c_\infty$ parameter spaces. Three phases can be identified:
stable SL, unstable SL and no SL. Stable SL only occurs in a tiny domain
of the parameter space which is in good quantitative agreement
with the UCLA SL experiments on argon bubbles \cite{bar95}.

For molecular gases  besides (i) shape instabilities and (ii) diffusive
instability also (iii) chemical instabilities have to be considered.
Then our hydrodynamic approach can be extended to gas mixtures such as air
as demonstrated in refs.\ \cite{loh96,bre96b}. Again, good agreement
with the UCLA experiments \cite{hil94,bar94,bar95,loe95,hil95}
is achieved.

We suggest to experimentally map out phase diagrams in the
$c_\infty/c_0 $ 
versus $P_a/P_0$ parameter space for various gas mixtures.
Detailed experimental data on the borderlines
between the stable, unstable, and no SL regimes will lead to  further 
refinements and improvements in our understanding of the dynamics 
of a sonoluminescing bubble.

The central question which cannot be answered within the present
approach is how {\it hot} the gas inside the bubble can become.
Progress on this point will require more sophisticated understanding of 
the gas dynamics {\it inside} the bubble.  Understanding how the gas 
temperature depends on experimental parameters such as forcing pressure,
gas concentration, 
or liquid temperature will allow for the creation of 
temperature controlled environments for chemical 
reactions within the bubble.

%For progress on this point
%detailed analysis of the gas dynamics will be necessary.
%If this has been better understood we may be close to the 
%{\it design} of bubbles with certain properties as the
%maximal temperature.
%Simply by controlling the external parameters as
%forcing pressure, gas concentration, or ambient liquid temperature
%one would thus be able to create {\it distinct} chemical microlabs.

\vspace{2cm}

\noindent
{\bf Acknowledgements:} We thank
K.\ Drese, T.\ Dupont, 
W.\ Eisenmenger, B.\ Gompf, S.\ Grossmann, B. Johnston,
D.\ Oxtoby, S.\ Putterman, and R.\ Rosales
for stimulating discussions
and in particular S.\ Grossmann for continuous encouragement. 
This work has been supported 
by the DFG through its SFB 185.
MB acknowledges a NSF postdoctoral 
fellowship, and support through the Sloan Research
Fund of the School of Science at MIT.

%\newpage

\vspace{2cm}

\begin{appendix}

\renewcommand{\theequation}{A.\arabic{equation}}
\section{Numerical scheme for the advection diffusion equation}\label{appa}
\setcounter{equation}{0}

\subsection{Transformation}
To numerically solve eq.\ (\ref{convdiff}) with the boundary
conditions (\ref{henry}) and (\ref{bc2}) we
first transform the independent space variable $r$.
Following \cite{vuo95}, we choose
\begineq
r=F(x,t) = \left( (R(t))^3 - R_u^3 \ln(1-x)\right)^{1/3}
\label{trafo}
\endeq
where $R_u$ is an adjustable length parameter.
All lengths are measured as multiples of $1\mu m$, all times in units
of $T=2\pi/\omega$, and all pressures in  atm.
After the transformation $r=F(x,t)$ the advection diffusion equation
(\ref{convdiff}) reads, if written as a local conservation law, 
\begineq
\partial_t (fc) +\partial_x (gc) - \partial_x(h\partial_xc) = 0,
\label{cd_transfer}
\endeq
where
\renewcommand{\arraystretch}{1.1}
\begineq
\left.
\begin{array}{lll}
f(x,t)&=& F^2 \partial_x F,  \\
g(x,t)&=& \dot R R^2 - F^2\partial_t F, \\
h(x,t)&=& D F^2/ \partial_x F.
\end{array}
\qquad\qquad\qquad
\right\}
\label{fgheq}
\endeq
The transformation (\ref{trafo}) is chosen to make $g(x,t)$
identically vanish and
thus to obtain a pure diffusion equation. We have
\begineq
\left.
\begin{array}{lll}
f(x,t)&=& R_u^3/(1-x), \\
g(x,t)&=& \,\,0, \\
h(x,t)&=& D (1-x)(F(x,t))^4/R_u^3.
\end{array}
\qquad \right\}
\label{fghseq}
\endeq
\renewcommand{\arraystretch}{1}
The time dependent range
$r\in [R(t),\infty ]$ has been mapped
to the constant range  $x\in [0,1]$.
The boundary conditions are Henry's law (\ref{henry}) which now reads
$c(x=0,t)= c_0 p(R,t)/P_0$
and $c(x=1,t)=c_\infty$.
The mass loss of the bubble, expressed in the new independent
variable $x$, is
\begineq
\dot{m} =  4 \pi   D R_u^{-3} R^4 \partial_x c|_{x=0}.
\label{gainlosss}
\endeq

\subsection{Discretization}
The interval $x\in [0,1]$ is discretized using a non-equidistant grid.
The grid must (i) provide sufficient volume for diffusion of gas outside
the bubble and (ii) sufficient resolution near the bubble radius for
a correct representation of gas concentration gradients.
We satisfy (i) by choosing $R_u\approx R_{max}$, where $R_{max}$ is the
maximum radius of a typical bubble oscillation. This ensures
$r^3(x\to1) \gg R^3$ even for relatively low resolution near $x=1$
(cf. eq. (\ref{trafo})). From this choice, we deduce
a criterion for the grid resolution $\delta x_R$ near the bubble surface
for a given resolution $\delta r_R$ in physical coordinates (e.g.
$\delta r_R= 0.01\mu\/m$). Expanding (\ref{trafo}) in $x\ll 1$, we
get
$$\delta x_R =\delta r_R {R_{min}^2\over R_{max}^3}.$$
This is a conservative estimate, requiring $\delta r_R$ to be resolved
even for the minimum bubble radius $R_{min}$. At times when the bubble
radius is larger, the resolution in $r$ will be even better with this
$\delta x_R$.

Typical bubble dynamics data
lead to $\delta x_R \approx 5\cdot10^{-6}$. Excellent resolution
at the bubble radius is needed for a correct representation of 
gas concentration gradients. It is, however, unnecessary at greater
distance from the bubble surface. Therefore, the grid density is
varied according to a power law to yield a fine grid near $x=0$ and
a coarse grid near $x=1$.

Overall,
we have $N$ grid points $x_1=0$, $x_2,\dots$, $x_{N-1}$, $x_N=1$.
The field $c(x,t)$ is defined on the grid, $c_i=c(x_i)$, and so are the
fields $f,g,h$. We define 
$dx_i=x_{i+1}-x_i$, $i=1,2,\dots N-1$ and
$dx_{a,i}=(dx_{i-1}+dx_i)/2$, $i=2,\dots N-1$, $dx_{a,1}=dx_1/2$,
$dx_{a,N}=dx_{N-1}/2$.

The discretization of eq.\ (\ref{cd_transfer}) has to guarantee
mass conservation,
\renewcommand{\arraystretch}{1.1}
\begineq
\left.
\begin{array}{lll}
%\begin{eqnarray}
0 &=& c^\theta_1-p(R^\theta(t))c_0/P_0,
\nonumber \\
0 &=& f_i{dc_i\over dt} - {1\over dx_{a,i}} \left(
{1\over 2}(h^\theta_i+h^\theta_{i+1})
 {c_{i+1}^\theta - c_i^\theta\over dx_i} - \right. \nonumber \\
  &-& \left.
{1\over 2}(h^\theta_{i-1}+h^\theta_{i})
{c_{i}^\theta - c_{i-1}^\theta\over dx_{i-1}}\right), \qquad i=2,\dots N-2,
\nonumber \\
0 &=& f_i{dc_i\over dt} - {1\over dx_{a,i}} \left(
-{1\over 2}(h^\theta_{i-1}+h^\theta_{i})
{c_{i}^\theta - c_{i-1}^\theta\over dx_{i-1}}
\right), \quad i= N-1,
\nonumber \\
0 &=& c^\theta_N-c_\infty.
%\label{discret}
%\end{eqnarray}
\end{array}
\qquad\qquad\qquad
\right\}
\label{discret}
\endeq
\renewcommand{\arraystretch}{1.0}
Here, $c_i^\theta$ are the concentrations
$c_i+\theta dc_i$ at time $t+\theta dt$
and $h_i^\theta = h_i(R^\theta)$ where $R^\theta$ is the radius
at time $t+\theta dt$. Correspondingly, $R_0^\theta$ is the ambient
radius at that time.
We choose $\theta=1$, i.e., a fully implicit method.
Equation (\ref{discret}) has to be assisted by
the Rayleigh-Plesset equation (\ref{rp}) and the proper discretization of
(\ref{gainlosss}) guaranteeing total mass conservation,
\begineq
0=
\rho_0 (R_0^\theta)^2 \dot R_0^\theta
- {h^\theta_1+h^\theta_2\over 2} {c_2^\theta - c_1^\theta \over dx_1}.
\label{disgain}
\endeq
We solve the  $N+2$ equations (\ref{discret}), (\ref{disgain}),
and (\ref{rp}) for the unknowns $dc_i$, $i=1,2,\dots N$,
$dR_0$, and $dR$ with Newton's method. The Jacobian is calculated
analytically. A time step control and adjustment is provided
by redoing every time
step $dt$ in two steps of width $dt/2$ each and then comparing the
result. 
For large forcing we need a very low tolerance of $10^{-5}$ per cent
to achieve sufficient numerical quality of mass conservation.
Note that this simulation covers time scales from picoseconds (for
good resolution of the bubble dynamics near the collapse) to tens of
milliseconds (for observation of diffusive growth or shrinking of the bubble).

\section{Adiabatic approximation and multiple equilibria}\label{appb}
\renewcommand{\theequation}{B.\arabic{equation}}
\setcounter{equation}{0}
In section \ref{compare} we showed that the adiabatic
approximation well describes
the full dynamics.
However, for (theoretical)
completeness we would like to caution in this appendix:
For the physical diffusion coefficient $D=2\cdot 10^{-9}m^2/s$ the 
simple adiabatic approximation may lead
to even {\it qualitatively} wrong results. E.g., take a bubble 
with $R_0=6.06\mu m$.
According to the adiabatic approximation one would expect that it
grows towards the
equilibrium
$R_0^{adiab}=6.097\mu m$ (see figure \ref{approach}).
However, it shrinks towards the physical equilibrium
$R_0^e=6.041\mu m$.
At the moment the experimental accuracy does not
allow to distinguish between these two sizes, but it may
improve one day. 
We can understand $R_0^e < R_0^{adiab}$ 
because the gas layer
around the bubble (figure \ref{profiles}) diffusively 
shrinks too strongly for
finite $D$, leading to a larger overall mass loss and 
thus smaller bubbles.
The order of magnitude of the $R_0^e$ shift can be estimated as
the width of the boundary layer:

\begineq
R_0^{adiab}-R_0^e \sim \delta_D=\sqrt{D\over\omega}
\propto D^{1/2} .
\label{dev}
\endeq
For the physical $D
=2\cdot 10^{-9} m^2/s$
we have
$R_0^{adiab}-R_0^e \approx 0.11 \mu m$
reproducing the numerical 
result $\approx 0.06\mu m$ quite accurately.
In the inset of fig.\ \ref{approach} we plot
$\log (R_0^{adiab}-R_0^e) $ vs $\log D$ and indeed find good agreement with
the scaling law (\ref{dev}).

Whether the deviations are considered to be serious or not 
depends on what is supposed to be analyzed.
If the focus lies on identifying equilibrium points, the
consequences can be quite drastic.
Note that for large enough $D$, the value of (\ref{equi2})
falls below the minimum values of
$\left<p(t)\right>_{t,4}/P_0$ (cf. figure \ref{p_tau}), 
thus making an
equilibrium solution impossible and leading to 
dissolution of the bubble. 

If the distance between  
subsequent equilibrium radii 
$\Delta R_0^{adiab,i} \equiv
R_0^{adiab,i}-R_0^{adiab,i-1}$
(where the index numbers label the equilibria in ascending $R_0$ order)
is supposed to be resolved, i.e., if one wants to find a 
one to one
correspondence between the adiabatic equilibria and the real 
ones, one has as to impose the following
condition on the approximation:
\begineq
\Delta R_0^{adiab,i} \gg {1\over 2}(R_0^{adiab,i}
- R_0^e) \sim {\delta_D\over 2}.
\label{condi1}
\endeq
Expressed in the Schmidt number $Sc$ it reads
\begineq
Sc \gg \left( {2 R_0 \over \Delta R_0^{adiab,i} }\right)^2 
\label{condi2}
\endeq
rather than simply $Sc \gg 1$ as one may naively expect.
With the correct values in the relevant parameter regime
$R_0 \approx 6\mu m$, $\Delta R_0^{adiab,i} \approx 
0.3\mu m$ we have $Sc \gg 1600$, which is only marginally fulfilled
by the physical $D$ corresponding to  $Sc\approx 3000$.

\end{appendix}

\vspace{1.5cm}

%\noindent

%\newpage
%\bibliography{sl_literatur}

\begin{thebibliography}{10}
\bibitem{szeri}
The recent calculations of Vuong and Szeri \cite{vuo95} 
incorporate dissipation mechanisms and do not find shocks.
\bibitem{boundary}
For a passive scalar as e.g.\ the gas  concentration
the analogous length scale is $\sqrt{D /\omega }$, where 
$D$ is the diffusion
constant. In section \ref{compare} we perform a full 
numerical simulation for the
diffusive problem and indeed see that $\sqrt{D/\omega }$ is the thickness of
the boundary layer around the bubble.
\end{thebibliography}

%\newpage

%\vspace{3cm}

%\centerline{\bf Tables}

%--------------------------------------------------------------------

\end{document}